\documentclass[preprint]{aastex631}

\begin{document}

\title{Exploration of the Near-Infrared Colors of Cold Y Dwarfs from the Ground and Space}

\correspondingauthor{Sandy Leggett}
\email{sandy.leggett@noirlab.edu}

\author[0000-0002-3681-2989]{S. K. Leggett}
\affiliation{Gemini Observatory/NSF's NOIRLab, 670 N. A'ohoku Place, Hilo, HI 96720, USA}

\author[0000-0001-6041-7092]{Mark W. Phillips}
\affiliation{University of Edinburgh, Royal Observatory, Blackford Hill, Edinburgh, EH9 3HJ, UK}

\author[0000-0001-6172-3403]{Pascal Tremblin}
\affiliation{Universite Paris-Saclay, UVSQ, CNRS, CEA, Maison de la Simulation, 91191, Gif-sur-Yvette, France}


\begin{abstract}

{\it JWST} has provided critical mid-infrared data for cold brown dwarfs. It has also provided low-resolution near-infrared spectra, and for faint sources these are the first spectra at these wavelengths. We use these data and other literature sources to synthesize near-infrared photometry on the MKO system for 19 T and Y dwarfs, on the  {\it Euclid} system for 44 T and Y dwarfs, and on the {\it Roman} systems for 48 T and Y dwarfs. We also synthesize  {\it Euclid} $I_E$ magnitudes for 15 T and Y dwarfs. Using the \citet{Beiler_2024} observational effective temperatures ($T_{\rm eff}$), together with ATMO 2020++ model colors, we show that the absolute $4.6~\mu$m magnitude can be used as a proxy for  $T_{\rm eff}$. We present a polynomial fit to the $M_{\rm W2}:T_{\rm eff}$ relationship for cool dwarfs with $250 \lesssim T_{\rm eff}$~K $\lesssim 1000$.
We select five Y dwarfs with $275 \lesssim T_{\rm eff}$~K $\lesssim 400$ which have a range in near- to mid-infrared colors. Comparison of the {\it JWST} spectral energy distribution to ATMO 2020++ models indicate that Y dwarfs which are bluer in $J -$ W2  are lower gravity or more metal-rich than their redder counterparts, with stronger CO and CO$_2$ absorption at $4.2 \lesssim \lambda~\mu$m $\lesssim 4.9$.  The near-infrared color diagrams show significant scatter, with complex dependencies on $T_{\rm eff}$, metallicity and gravity. 
In order to disentangle these effects, opacity sources for cool model atmospheres need to be more complete at $\lambda < 1~\mu$m.
\end{abstract}



\bigskip
\section{Introduction} 

The James Webb Space Telescope ({\it JWST}) has provided fundamentally important mid-infrared spectroscopy and photometry for the coldest known bodies outside our solar system, the brown dwarfs \citep[e.g.][]{Barrado_2023, Beiler_2023,
Calissendorff_2023, Beiler_2024, Faherty_2024, Luhman_2024, Tu_2024, Leggett_2025}. Brown dwarfs have insufficient mass for sustained energy production by nuclear fusion, and cool over periods of gigayears \citep[e.g.][their Figure 13]{Marley_2021}. The coldest brown dwarfs are classified as Y dwarfs, and have effective temperatures ($T_{\rm eff}$) lower than 500~K \citep{Cushing_2011, Kirkpatrick_2021a}. The warmer T-type brown dwarfs span a range in temperature of $ 500 \lesssim T_{\rm eff}$~K $\lesssim 1200$ \citep{Kirkpatrick_2021a}.

More than 90\% of the total energy emitted by a Y dwarf emerges at wavelengths $\lambda > 3~\mu$m \citep[e.g.][their Figure 1]{Leggett_2025}. However the near-infrared regime is the region most readily accessible by ground-based telescopes, and is the region being explored by the recently launched or soon to launch (at the time of writing) space missions {\it Euclid} \citep{Euclid_2022} and {\it Roman} \citep{Roman_2024}. {\it JWST}'s NIRSpec prism \citep{NIRSpec_2022} has also provided some of the first near-infrared spectra of the faintest Y dwarfs.

In this work we use these {\it JWST} spectra to synthesize MKO-system $YJHK$ magnitudes \citep{MKO_2002} for late-T and Y dwarfs, to supplement the existing MKO photometry set for brown dwarfs. We also synthesize {\it Euclid} \citep{Euclid_NISP, Euclid_VIS} and {\it Roman} \citep{Roman_2024} photometry, to extend the work done by \citet{Sanghi_2024} to cooler objects, using {\it JWST} spectra and other literature sources. 

Section 2 illustrates the wavelength regions sampled by the MKO, {\it Euclid} and {\it Roman} near-infrared filters. Section 3 
presents the MKO photometry. Using the updated photometry, we select a sample of five of the coldest Y dwarfs, which have a range of near- to mid-infrared colors, and estimate the properties of these objects in Section 4. 
We use synthetic spectra generated by the ATMO 2020++ model suite for the analysis \citep{Phillips_2020, Leggett_2021, Meisner_2023, Leggett_2025}. These models use a semi-empirical pressure-temperature relationship for the atmosphere \citep{Leggett_2021, Leggett_2025}. Although imperfect, as we illustrate below, at this time they reproduce observations of late-T and Y dwarfs better than other available forward (grid) models \citep[e.g.][their Figure 1]{Albert_2025}.
In Section 5 we present synthetic
{\it Euclid} photometry and illustrate the location of cold brown dwarfs in color diagrams. In Section 6 we do the same with {\it Roman} photometry. Section 7 gives our Conclusions.

\bigskip
\section{MKO, Euclid and Roman Near-Infrared Filter Bandpasses}


\begin{figure}
\includegraphics[angle=-90, width = 7.5 in]
{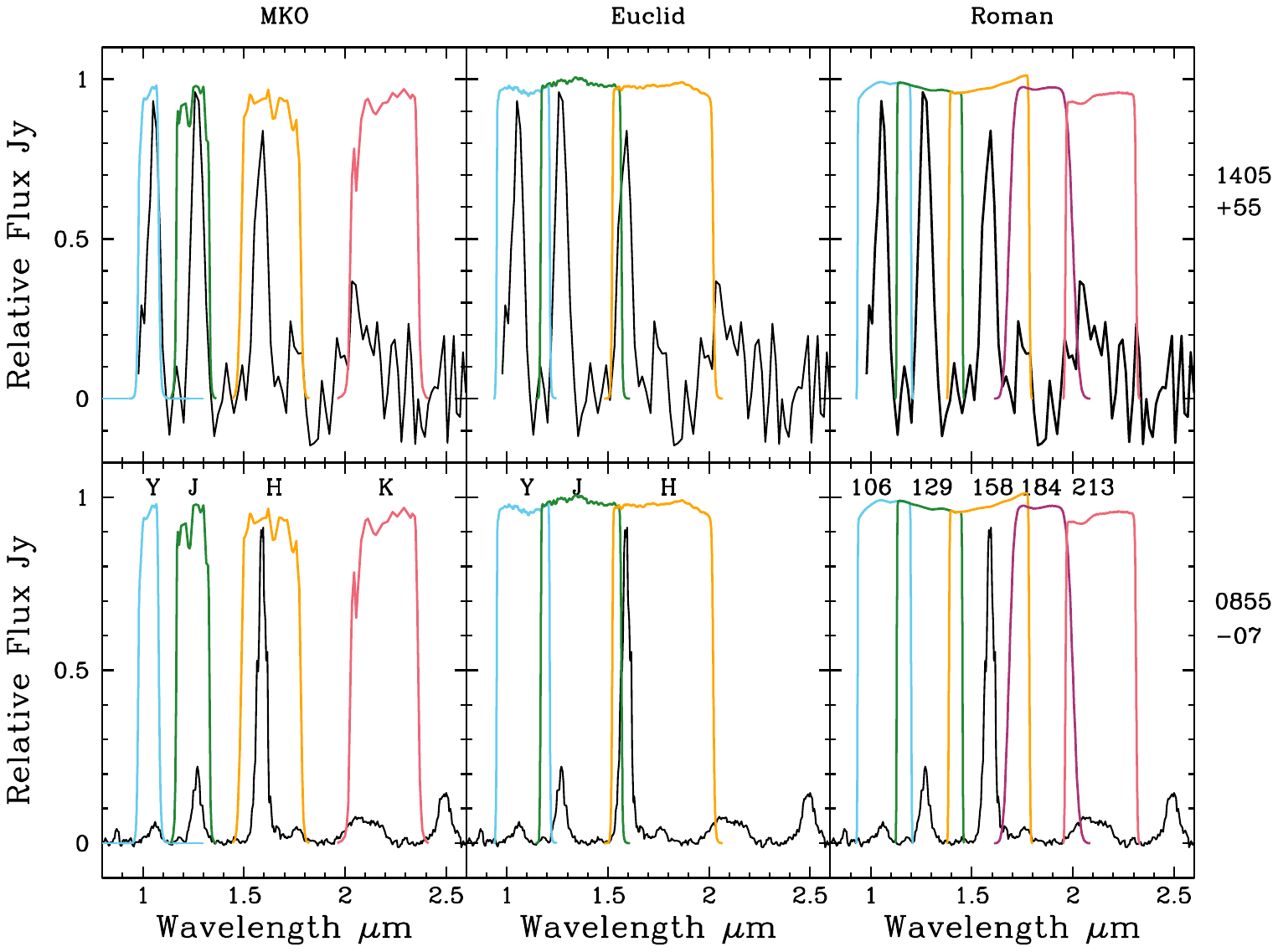}
\caption{Black lines are {\it JWST} spectra for 
WISE  140518.40$+$553421.4 (1405$+$55; upper three panels, from \citet{Beiler_2024}) and WISE 085510.83$-$071442.5 (0855$-$07; lower three panels, from \citet{Luhman_2024}). 
1405$+$55 has a luminosity-based $T_{\rm eff}$ of 392~K \citep{Beiler_2024}, and       0855$-$07 has a lower $T_{\rm eff}$ -- between 240~K and 299~K -- based on forward grid model and retrieval analyses \citep{Kuhnle_2024}. Colored lines show the filter profiles for the MKO, {\it Euclid} and {\it Roman} filters, from left to right. Note that, in order to find other examples of the so far unique 0855$-$07, sensitivity in the $H$-band is required.
}
\end{figure}

Figure 1 shows the bandpasses for the MKO $Y, J, H,$ and $K$ filters; the {\it Euclid} $Y_E, J_E,$ and $H_E$ used by the NISP instrument; and the {\it Roman} F106, F129, F158, F184, and F213 bandpasses. The {\it Roman} filters F062 and F087 sample regions where there is negligible signal from late-T or Y dwarfs \citep[e.g.][and Figure 1]{Leggett_2012} and are not included in this work. The {\it Roman} F184 filter samples a region of strong absorption (Figure 1) and is not ideal for brown dwarf work; however the {\it Roman} High Latitude Imaging Survey as specified by
\citet{Troxel_2023}
uses the F106, F129, F158 and F184 filters  and so we include it here. The F213 filter is more suitable for brown dwarf studies than the F184, as we show in Section 6.

The MKO filters were designed to avoid wavelengths where the Earth's atmosphere absorbs light, which coincides with absorption bands in cold brown dwarf atmospheres, hence the MKO filters nicely sample the brown dwarf flux peaks. The {\it Roman} filter set also samples the flux peaks cleanly (except for F184), while the broader {\it Euclid} filters include a contribution from the redder flux peak at the red end of each of the three near-infrared filter bandpasses. 

{\it Euclid} also images the sky at optical wavelengths using the VIS camera and the $I_E$ filter, which spans 540~nm to 930~nm \citep{Euclid_VIS}. With a Wide Survey limit of $I_E \approx 26_{AB}$ \citep{Euclid_WideSurvey} Y0 dwarfs will have to be closer than about 10~pc to be detected. Nevertheless, a limit on the the optical to near-infrared colors is useful for differentiating between high redshift sources and cold brown dwarfs. We return to this topic in Section 5.  

\bigskip
\section{New Synthetic MKO Photometry}

Table 1 gives synthetic MKO-system $YJHK$ Vega magnitudes. For all but one object (WISE 075108.79$-$763449.6), the magnitudes were calculated using {\it JWST} low-resolution NIRSpec prism spectra, which cover the wavelength range of  0.6$\mu$m to 5.3$\mu$m  \citep{NIRSpec_2022}. For all but three of these the {\it JWST} data were obtained as part of project GO 2302, PI M. Cushing, and published in \citet{Beiler_2023, Beiler_2024}. For WISE 03480772$-$6022270 and WISE 182831.08$+$265037.8 (hereafter 1828$+$26) the data were obtained as part of the GTO project 1189 with PI T. Roellig; these data are published in \citet{Tu_2024} and \citet{Leggett_2025} respectively. For WISE 085510.83$-$071442.5 (hereafter 0855$-$07)  the data were obtained as part of the  GTO program 1230, PI C. Alves de Oliveira,
and published in \citet{Luhman_2024}. 

The uncertainties in the synthetic photometry determined from the NIRSPEC data were estimated from the noise in each spectrum across the filter bandpasses. We adopt a lower limit on the uncertainty of 5\%, as described in \href{https://jwst-docs.stsci.edu/jwst-calibration-status/nirspec-calibration-status/nirspec-fixed-slit-calibration-status#gsc.tab=0}{
the NIRSpec photometric calibration documents}.

For WISE 075108.79$-$763449.6 \citep[COCONUTS-2b]{Zhang_2021a}, the Gemini spectrum published by \citet{Zhang_2025_COCO} 
was used to calculate the photometry, supplemented by the lower resolution and lower signal-to-noise ratio (SNR) Magellan spectrum published by \citet{Kirkpatrick_2011}. 
We used the Magellan spectrum to extend the Gemini spectrum to shorter wavelengths in order to synthesize $Y$-band photometry. We also used it to explore the uncertainties in the derived colors. For both the Gemini and the Magellan spectrum the SNR across
the $K$-band is low,  and we estimated the uncertainty in the $K$ magnitude from the range of values found applying different clipping and smoothing techniques to the data. The colors were synthesized from the spectra, and the absolute flux was set using the $J_{MKO}$ value determined by \citet{Kirkpatrick_2011}.


\begin{deluxetable*}{lrrrrcrrrr}
\setlength{\tabcolsep}{2pt}
\tabletypesize{\scriptsize}
\tablecaption{New Synthesized MKO Near-Infrared Photometry, Vega mags}
\tablehead{\\[0.01in]
\colhead{AllWISE Name} & 
\multicolumn{4}{c}{Previous Measurement} & \colhead{Ref.} & 
\multicolumn{4}{c}{This Work}\\
\colhead{} &
\colhead{$Y$} & 
\colhead{$J$} &
\colhead{$H$} & 
\colhead{$K$} &
\colhead{} &
\colhead{$Y$} & 
\colhead{$J$} &
\colhead{$H$} & 
\colhead{$K$}}
\startdata
024714.52$+$372523.5	 &  & 18.00 $\pm$ 0.06 & 18.24 $\pm$ 0.19 &  & Ki19,Be20 & 19.18 $\pm$	  0.05  & 	18.01	$\pm$  0.05	 & 18.34	$\pm$  0.05  & 	18.61	$\pm$   0.05 \\
031325.96$+$780744.2	 &  & 17.67 $\pm$ 0.07 & 17.67 $\pm$ 0.07 &  & Be14,Le19 &
18.31 $\pm$	  0.05  & 	17.46 $\pm$	  0.05 	 & 17.77	$\pm$   0.05 	 & 18.06	$\pm$   0.05 \\
034807.72$-$602227.0 &  & 15.00 $\pm$	0.03 & 15.49 $\pm$	0.10 & 15.60 $\pm$	0.10 & 2M,VHS &
16.06 	$\pm$   0.05  & 15.00 	$\pm$   0.05  & 15.36 	$\pm$   0.05  & 15.48	$\pm$    0.05 \\
035934.06$-$540154.6	 & 21.84 $\pm$  0.11 & 21.53 $\pm$  0.11 & 21.72 $\pm$  0.17 & 22.80 $\pm$  0.30 & Le15 
& 21.76		$\pm$    0.09 	 & 21.42		$\pm$   0.07 	 & 21.62		$\pm$   0.08 	 & 21.69	$\pm$ 	  0.16 \\
043052.92$+$463331.6	 &   & 19.06 $\pm$  0.02 & 19.24 $\pm$  0.12 &  &  Ki19,Be20 &
19.68		$\pm$    0.05 	 & 19.02	$\pm$ 	  0.05  & 	19.18		$\pm$   0.05 	 & 20.39	$\pm$ 	0.20\\
053516.80$-$750024.9	 & 22.73 $\pm$ 0.30 & 22.50 $\pm$ 0.30 & 23.34 $\pm$ 0.34 &  & Le15 &
22.70		$\pm$   0.11 	 & 22.44	$\pm$ 	  0.09  & 	22.33		$\pm$   0.13 	&	\\
073444.02$-$715744.0	 & 
21.02 $\pm$ 	0.05 & 20.28 $\pm$ 	0.23 & 20.92 $\pm$ 	0.12 & 20.96 $\pm$ 	0.15 & Le15 &
20.87	$\pm$ 	  0.05 	 & 20.33		$\pm$   0.05 	 & 20.70		$\pm$   0.05 	 & 21.13	$\pm$ 	  0.11 \\
075108.79$-$763449.6$^*$ & 20.02 $\pm$ 0.15 &  19.34 $\pm$  0.05 & 19.68 $\pm$ 0.13 & 20.03 $\pm$ 0.20 &  Ki11,Le15 &  19.95 $\pm$   0.07  &  19.34 $\pm$   0.05  &  19.59 $\pm$   0.07  &  19.15 $\pm$  0.20 \\
082507.35$+$280548.5	 & 22.66 $\pm$ 	0.20 & 22.53 $\pm$ 	0.20 & 23.09 $\pm$ 	0.18 &  & Le17 &
22.37	$\pm$ 	  0.16 	 & 22.66	$\pm$ 	   0.13  & 	22.91	$\pm$ 	0.20	 & 22.53 $\pm$ 	0.30\\
085510.83$-$071442.5 &
26.54 $\pm$ 	0.21 & 25.45 $\pm$ 	0.24 & 23.83 $\pm$ 	0.25 &  
& Le17,Le19 &  26.30	$\pm$   0.06  & 25.03 $\pm$   0.05  & 23.57	$\pm$   0.05  & 24.91 $\pm$   0.07 \\
104756.81$+$545741.6 & 	
 &  &  &  & &
21.54 $\pm$ 	  0.05 	 & 21.22 $\pm$ 	   0.05 	 & 21.72 $\pm$ 	  0.05  & 	21.74	$\pm$   0.07 \\
120604.38$+$840110.6 & 	
20.89 $\pm$ 0.10 & 20.38 $\pm$ 0.10 & 20.97 $\pm$ 0.12 &  & Le17 &
20.95 $\pm$ 	  0.05 	 & 20.38 $\pm$ 	  0.05 	 & 20.88	$\pm$   0.05 	 & 20.77	$\pm$   0.07 \\
144606.62 $-$231717.8 & 	& 23.20 $\pm$ 	0.14 & &  & Le21 & 
23.30	$\pm$   0.09  & 	22.92	$\pm$    0.09 	 & 23.19 $\pm$ 	  0.11 	 & 22.92 $\pm$ 	  0.16 \\
150115.92$-$400418.4	 &  & 16.10 $\pm$ 	0.01 & 16.38 $\pm$ 	0.04 & 16.31 $\pm$ 	0.05 & Ki19,VHS &
17.29	$\pm$   0.05 	 & 16.15 $\pm$ 	  0.05 	 & 16.45 $\pm$ 	  0.05  &	16.44 $\pm$ 	  0.05 \\
154151.66$-$225025.2	 & 21.46 $\pm$ 	0.13 & 21.12 $\pm$ 	0.06 & 21.07 $\pm$ 	0.07 & 21.70 $\pm$ 	0.20 & Le13,Le15 &
21.39 $\pm$ 	  0.09 	 & 21.25 $\pm$ 	  0.07 	 & 21.68 $\pm$ 	  0.16  &	21.78 $\pm$ 	0.25\\
182831.08$+$265037.8	&	23.03  $\pm$ 0.17  &	23.48 $\pm$ 0.30    &	22.73  $\pm$  0.13 &	23.48 $\pm$ 0.36 & Le13,Le15 
& 23.00  $\pm$	  0.05 	 &  23.05  $\pm$ 	  0.05 	  &   22.59  $\pm$ 	  0.05 	& 22.94  $\pm$ 	  0.05 	\\
210200.15$-$442919.5	 &  & 18.30 $\pm$ 	0.04 & 18.58 $\pm$ 	0.07 & 18.99 $\pm$ 	0.26 & Ki19,VHS 
& 19.02 $\pm$ 	  0.05  & 	18.22 $\pm$ 	  0.05 	 & 18.55 $\pm$ 	  0.05 	 & 18.64 $\pm$   0.05 \\
215949.54$-$480855.2 	 &   & 18.84  $\pm$ 0.10 & 19.25  $\pm$ 0.12 &  &   Ki19,VHS &
19.78 $\pm$ 	  0.05 	 & 18.88 $\pm$ 	  0.05 	 & 19.19 $\pm$ 	  0.05 	& 20.09 $\pm$ 	  0.07 \\
235402.79$+$024014.1  & 	& 22.72  $\pm$  0.13 & 22.53  $\pm$ 0.28 & & Le17 &
22.45	 $\pm$   0.11 	 & 22.92 $\pm$ 	  0.11 	 & 22.98	 $\pm$   0.16 &		\\
\enddata
\tablecomments{All synthesized photometry uses {\it JWST} spectra except for WISE 075108.79 $-$763449.6, see Section 3. For this brown dwarf the spectrum was calibrated using the previously published $J$ magnitude.}
\tablenotetext{*}{Also known as COCONUTS-2b or L 34-26b.}
\tablerefs{
2M -- 2MASS, \citet{Skrutskie_2006};
Be14 -- \citet{Beichman_2014};
Be20 -- \citet{Best_2020};
Ki11 -- \citet{Kirkpatrick_2011};
Ki19 -- \citet{Kirkpatrick_2019};
Le13 -- \citet{Leggett_2013};
Le15 -- \citet{Leggett_2015};
Le17 -- \citet{Leggett_2017};
Le19 -- \citet{Leggett_2019};
Le21 -- \citet{Leggett_2021};
VHS -- VISTA VHS, \citet{Sutherland_2015}.
}
\end{deluxetable*}

In addition, we estimated the MKO $J$ magnitude for the Y dwarf WISE 023842.60$-$133210.7 from the {\it Hubble Space Telescope} ({\it HST}) F110W image. The  {\it HST} data were obtained as part of program 16243 (PI: F. Marocco). The data were calibrated using the zeropoint provided in the image header, and the transformation from F110W to MKO $J$ was determined using the relationship shown in \citet[][their Figure 6]{Meisner_2023}, adopting a spectral type of Y1. We find $J = 22.02 \pm 0.25$ for WISE 023842.60$-$133210.7.

The new synthesized values are within 3$\sigma$ of the previous measurements, except in two cases. Possible explanations for the discrepancies include underestimated uncertainties in the data, or source variability \citep[e.g.][]{Buenzli_2014}. The $H$ magnitudes for WISE 154151.66$-$225025.2 (1541$-$22) differ by $3.5\sigma$. The  {\it JWST} spectrum gives $H = 21.68 \pm 0.16$, compared to the previous measurement of $21.07 \pm 0.07$ from \citet{Leggett_2015}. Because the NIRSpec prism covers $YJHK$ simultaneously the synthesized colors will be robust, and we adopt the synthesized values in this work. Also, the $K$ magnitudes for COCONUTS-2b differ by $3.1\sigma$.  We find $K = 19.15 \pm 0.20$, compared to the previous measurement of $20.03 \pm 0.20$ from \citet{Leggett_2015}. The Gemini FLAMINGOS-2 $YJHK$ spectrum was obtained in two overlapping wavelengths settings, and so we adopt the synthesized values here which should provide consistent colors.

\bigskip
\section{A Sample of Five Y Dwarfs and their Properties}

\subsection{Observational Data and Model Sequences}

Figure 2 shows the absolute magnitude in the {\it WISE}
$4.6~\mu$m bandpass W2 \citep{Wright_2010} 
and the $J - H$ near-infrared color, 
as a function of the near- to mid-infrared $J -$ W2 color, for late-T and Y dwarfs. The photometry is taken from the compilation presented in the Appendix, which supercedes that given in
\citet{Leggett_2025}. The new photometry in Table 1 is included, and we supplemented the database 
by including $YJHK$ values from Table 1 of \citet{Best_2024}, adding sources and/or data previously omitted. We also added metal-poor brown dwarfs from the ``Primeval'' series of papers by Zhang et al. \citep[e.g.][]{Zhang_2019}, in order to broaden the metallicity range of the sample. 
Synthesized near-infrared photometry for the extremely metal-poor brown dwarf WISE 153429.75$-$104303.3 (hereafter 1534$-$10) from Faherty et al. (2025, Nature in press) was included. Parallaxes for six Y dwarfs were updated using the values given by \citet{Fontanive_2025}\footnote{WISE 041022.71$+$150248.5, 071322.55$-$291751.9, 073444.02$-$715744.0, 173835.53$+$273258.9, 205628.90$+$145953.3, 222055.31$-$362817.4}.
The updated photometry database is provided in the Appendix.

\begin{figure}
\includegraphics[angle=-90, width = 7.5 in]
{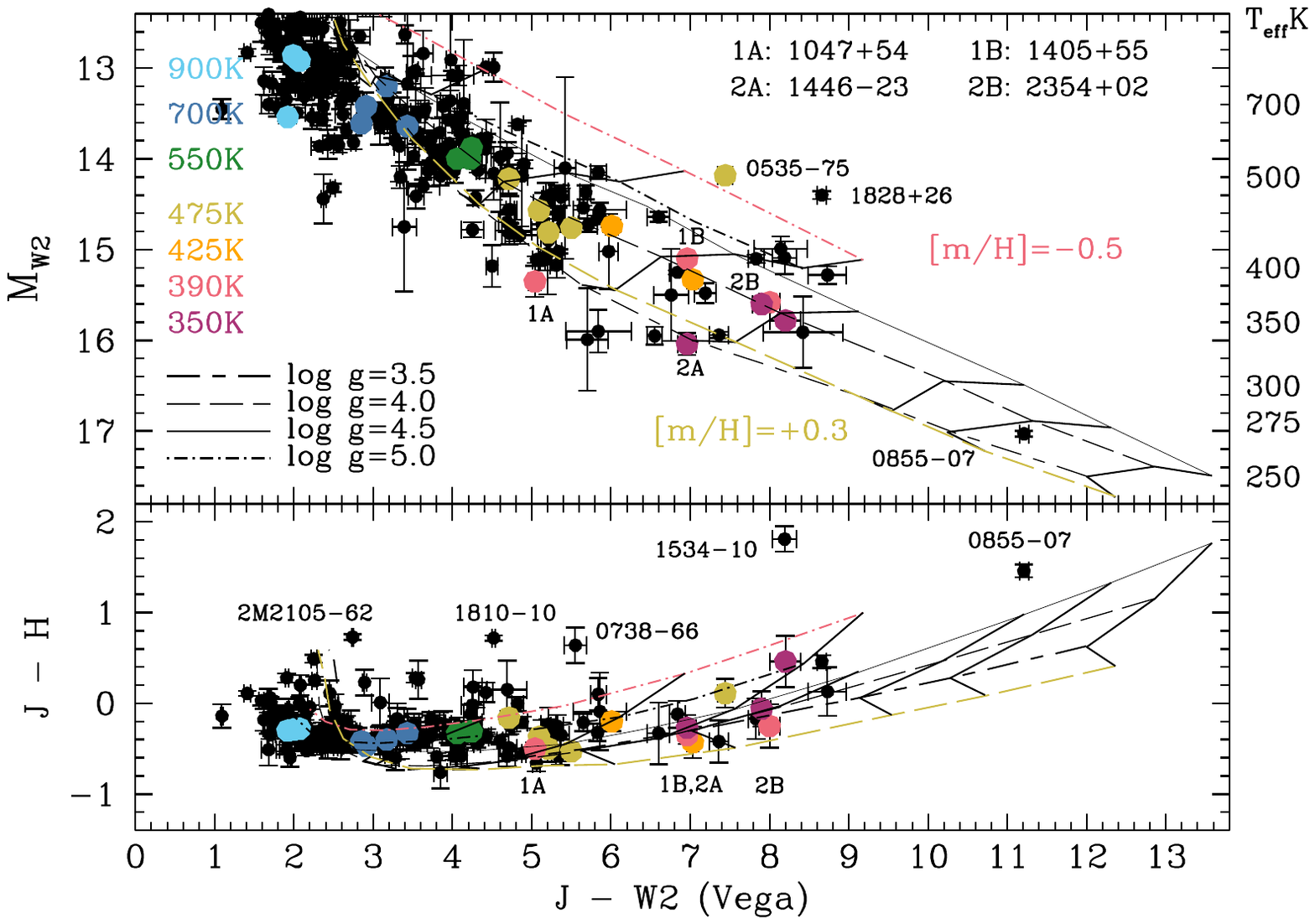}
\vskip -0.1in
\caption{
Color-magnitude and color-color diagrams for the sample of late-T and Y dwarfs presented in the Appendix. 
The lower panel excludes intrinsically bright objects with $M_{\rm W2} < 12.4$ to facilitate direct comparison of the upper and lower panels.
Colored symbols are brown dwarfs in the \citet{Beiler_2024} sample, where the color indicates the Beiler et al. $T_{\rm eff}$ value, shown along the upper left axis. Lines are ATMO 2020++ sequences \citep{Leggett_2021, Meisner_2023, Leggett_2025} for different gravities and metallicities as indicated in the legends. The log $g = 5.0$ sequences terminate at $T_{\rm eff} = 400$~K as 
evolutionary models  calculate that colder high-gravity objects would be older than 15~Gyr \citep[see Figure 3]{Marley_2021}.
Solid lines across the sequences are isotherms with $T_{\rm eff}$ values indicated on the upper right axis. The Y dwarfs indicated as 1A and 1B are 400~K objects which we compare to the 350~K Y dwarfs 2A and 2B in Section 4. The cold WISE 0855$-$07 is identified.
In the upper panel the likely binaries
WISE 053516.80$-$750024.9 and 1828$+$26 are identified \citep{Leggett_2024, Leggett_2025}. In the lower panel these four objects with very red $J - H$ colors are identified: 2MASS 21052823$-$6235461 \citep{Luhman_PM_2014}, 
WISE 073844.52$-$664334.6 \citep{Meisner_2021},
WISE 1534$-$10 \citep{Kirkpatrick_2021b}, and WISE 181006.00$-$101001.1 \citep{Schneider_2020}. WISE 1534$-$10 and 1810$-$10 are extremely metal-poor with 
[m/H] $\approx -2.2$ and $-1.5$ dex respectively 
\citep[Faherty et al. 2025, Nature in press and][]{Zhang_2025_1810}.
}
\end{figure}

\begin{figure}
\vskip -0.8in
\hskip 0.5in
\includegraphics[angle=0, width = 5.5 in]
{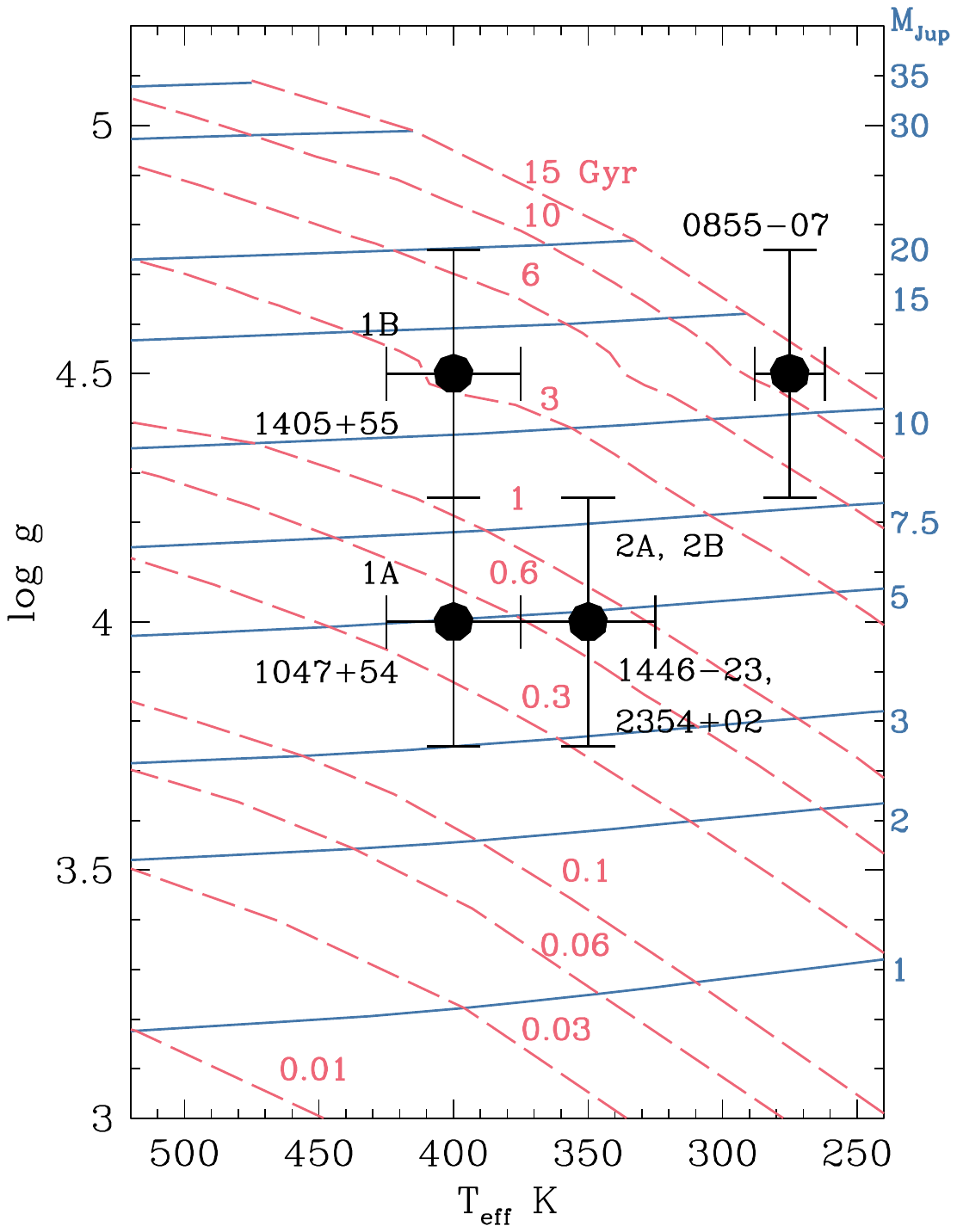}
\vskip -0.4in
\caption{Evolutionary tracks from solar-metallicity  models by \citet{Marley_2021}. Axes are atmospheric effective temperature and surface gravity expressed as $g$ cm~s$^{-2}$.  Blue solid lines are tracks of constant mass, with mass in Jupiter mass units shown along the right axis. Red dashed lines are lines of constant age, with age shown in Gyr above each line.  \citet{Marley_2021} find that a metal-rich brown dwarf with [m/H] $= +0.5$ and a mass $\sim~10~M_{Jupiter}$ will be  $\sim 10$\% older for a given $T_{\rm eff}$ or $\sim 25$~K warmer for a given age.  The filled circles represent our estimates of the properties of the five brown dwarfs listed in Table 2. 
We find that both 1047$+$54 and 1446$-$23 are  metal-rich, and so they are likely to be slightly older than the tracks imply.
}
\end{figure}

The ATMO 2020++ sequences \citep{Phillips_2020, Leggett_2021, Meisner_2023, Leggett_2025} shown in Figure 2 have metallicities of $-0.5$, 0.0, and $+0.3$ dex, and log $g$ values of 3.5, 4.0, 4.5, and 5.0 dex. The metallicity range is typical of the solar neighborhood \citep[e.g.][their Figure 4]{Hinkel_2014}, and 
the gravities correspond to a large age range of 10 Myr to 15 Gyr
according to evolutionary models \citep[for these temperatures,][]{Marley_2021}, encompassing the ages of young moving groups as well as older solar neighbourhood stars \citep[e.g.][]{Bell_2015, Haywood_2013}. Figure 3 illustrates the evolutionary tracks from \citet{Marley_2021} for the temperatures and gravities considered here.

Throughout this paper we show ATMO 2020++ single-gravity sequences. Atmospheric iso-gravity sequences are very nearly iso-mass sequences thanks to the small changes in radius during brown dwarf evolution \citep[e.g.][their Figure 16]{Marley_2021}. This can also be seen in Figure 3. Temperature then becomes a measure of age as the brown dwarf cools (Figure 3). The log $g =$ 3.5, 4.0, 4.5, and 5.0 sequences in Figure 2 (and elsewhere in this paper) therefore correspond approximately to cooling tracks for 2, 5, 13, and 30 Jupiter-mass brown dwarfs.

\subsection{Trends with Temperature}

In Figure 2, brown dwarfs with $T_{\rm eff}$ values determined by \citet{Beiler_2024} are shown as colored dots. \citet{Beiler_2024} use {\it JWST} spectral energy distributions (SEDs) to calculate bolometric luminosities which, when combined with radii estimated from evolutionary models, gives $T_{\rm eff}$ via Stefan's Law. The symbol color in Figure 2 indicates $T_{\rm eff}$. 
The trend of decreasing  $T_{\rm eff}$ with increasing $M_{\rm W2}$ is apparent, and the ATMO 2020++ sequences 
calculate $T_{\rm eff}$ values which agree well with the empirical values. Both the empirical and model calculations show  a small range in $M_{W2}$ of a few tenths of a magnitude for a given $T_{\rm eff}$, making $M_{W2}$ (or similar 4.6~$\mu$m bandpass) a useful indicator of temperature. 

Figure 4 shows the \citet{Beiler_2024} individual 
$T_{\rm eff}$ values from their Table 4, with the 
associated $M_{W2}$ values from our data compilation. Also shown are the $M_{W2}$:$T_{\rm eff}$ values from the ATMO 2020++ PH$_3$-free models, averaged over the metallicity and gravity sequences in Figure 2 (as shown on the right axis of Figure 2).
We fit a weighted polynomial to the Beiler et al. data points, adding the three coldest ATMO 2020++ data points to extend the polynomial to colder values. An error of $\pm~20$~K was adopted for the model values, which is representative of the spread with metallicity and gravity (Figure 2). The $rms$ deviation of the fit is 21~K, excluding two outliers (the T6 WISE 150115.92$-$400418.4, and the T7 WISE 201824.97$-$742327.5). The relationship can be used to estimate $T_{\rm eff}$ but it will not be representative of brown dwarfs which are multiple, or have atypical gravities and metallicities. We find:
$$ y= 2.58091e+05 - 6.40361e+04*x + 5.97992e+03*x^2 - 2.4849e+02*x^3 + 3.87238*x^4 $$
where $y = T_{\rm eff}$ and $x = M_{W2}$.

\begin{figure}
\hskip 0.5in
\includegraphics[angle=-90, width = 6.5 in]
{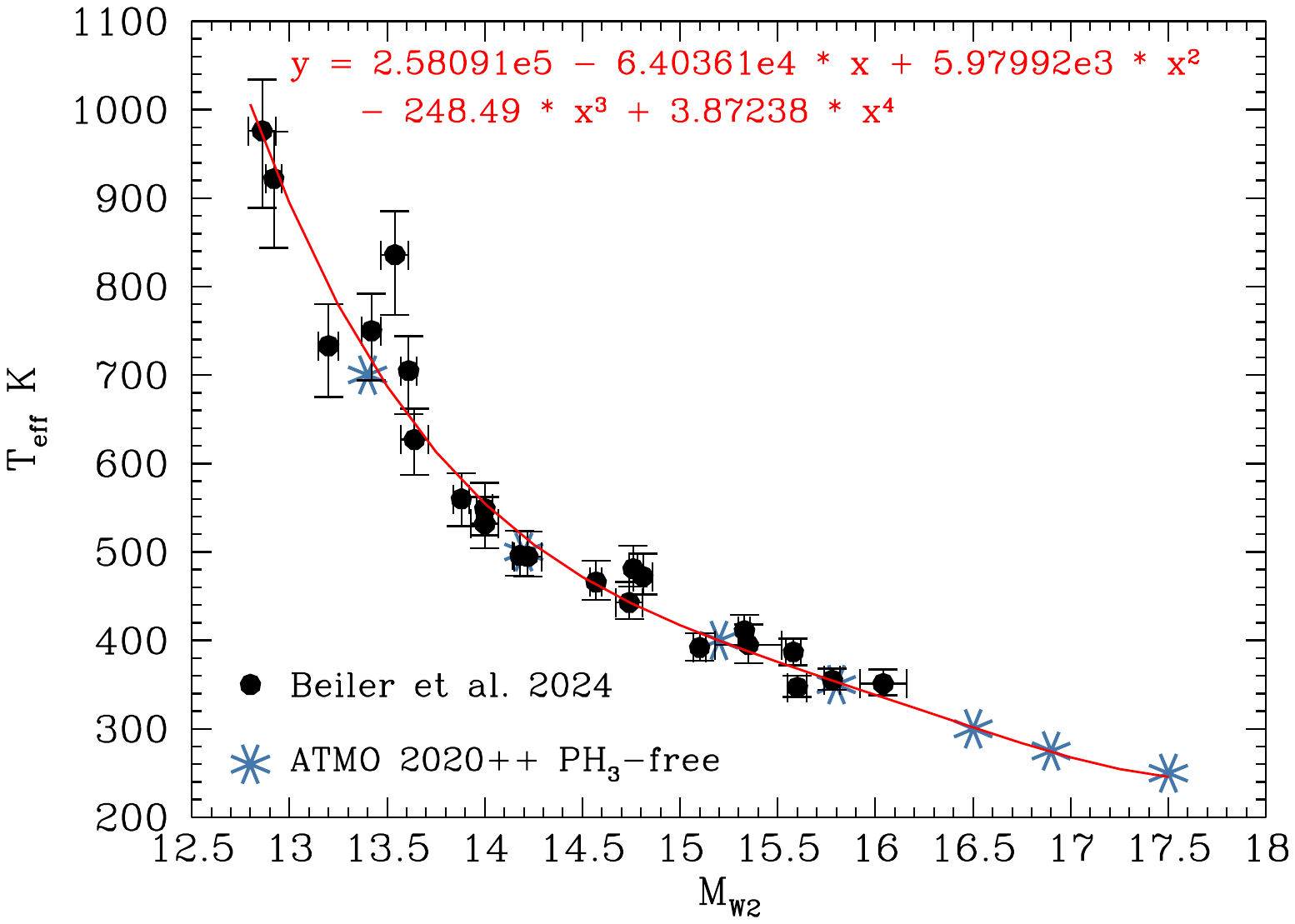}
\vskip -0.2in
\caption{Empirical relationship between  $T_{\rm eff}$ and  $M_{W2}$, combining the \citet{Beiler_2024} luminosity-based $T_{\rm eff}$ values and the ATMO 2020++ PH$_3$-free synthetic photometry (upper right axis of Figure 2). The $rms$ deviation, excluding two outliers, is 21~K.}
\end{figure}

\subsection{Five Y dwarfs for Spectral Analysis}

For this work we compare the properties of two Y dwarfs with $T_{\rm eff} \approx 400$~K and two with $T_{\rm eff} \approx 350$~K, each pair differing in $J -$ W2 color (Figure 2).  We add the 275~K Y dwarf 0855$-$07 to the analysis.
The 400~K Y dwarfs are WISE 104756.81$+$545741.6 (1047$+$54) and 140518.40$+$553421.4 (1405$+$55) with $J -$ W2 values of 5.04 and 6.96 respectively. The 350~K Y dwarfs are WISE 144606.62$-$231717.8 (1446$-$23) and 235402.79$+$024014.1 (2354$+$02) with $J -$ W2 values of 6.96 and 7.90 respectively. The ATMO 2020++ models indicate that the bluer dwarfs are more metal-rich and/or lower gravity than the redder dwarfs. We explore this further below.  Table 2 gives estimated properties of 
the five Y dwarfs from the literature and from this work.

\begin{deluxetable*}{crrrrcrrrrr}
\vskip 0.2in
\tabletypesize{\footnotesize}
\tablecaption{Estimated Properties of Five Y Dwarfs}
\tablehead{\\[0.01in]
\colhead{Short} & 
\colhead{Beiler\tablenotemark{a}} & 
\multicolumn{3}{c}{Other\tablenotemark{b}} & \colhead{Other Ref.\tablenotemark{b}} & 
\multicolumn{3}{c}{This Work\tablenotemark{c}} & Mass\tablenotemark{d} & Age\tablenotemark{d} \\
\colhead{Name} & 
\colhead{$T_{\rm eff}$ K} & 
\colhead{$T_{\rm eff}$ K} & 
\colhead{log $g$} & 
\colhead{[m/H]} &  &
\colhead{$T_{\rm eff}$ K} & 
\colhead{log $g$} & 
\colhead{[m/H]} & $M_{\rm Jup}$ & Gyr 
}
\startdata
0855$-$07 &  & 298 & 4.8 &  & K-retrieval & 275 $\pm 13$ & 4.5 $\pm 0.25$ & 0.0 $\pm 0.2$ & 12$^{+3}_{-4}$ & 12$^{+3}_{-7}$ \\
  &  & 297 & 4.2 & 0.0 & K-ATMO2020$++$ &  &  &  &  &  \\
    &  & 250 & 3.5 & $+0.5$ & K-L-non-eq.-clear &  &  &  &  &  \\
     &  & 275  & 3.2 & $+0.4$ & K-Elf-Owl &  &  &  &  &  \\
1047$+$54& 395$^{+23}_{-21}$ & 381 & 2.5 & $-1.0$ & Tu-ATMO2020$++$ & 400 $\pm 25$ & 4.0 $\pm 0.25$ & $+0.3 \pm 0.2$ & 5$^{+3}_{-2}$ & 0.5$^{+0.5}_{-0.3}$ \\
1A &  & 428 & 3.7 & +0.3 & Tu-Elf-Owl & &  &  &  &  \\
1405$+$55 & 392$^{+16}_{-15}$ & 402 & 4.6 & 0.0 & Tu-ATMO2020$++$ & 400 $\pm 25$ & 4.5 $\pm 0.25$ & 0.0 $\pm 0.2$ & 13$^{+7}_{-5}$ & 3$^{+4}_{-2}$ \\
 1B      &     &   389 & 3.3 & 0.0 & Tu-Elf-Owl & &  &  &  &  \\
1446$-$23 & 351$^{+16}_{-13}$ &  363 & 3.9 & 0.0 & Tu-ATMO2020$++$ & 350 $\pm 25$ & 4.0 $\pm 0.25$ & $+0.3 \pm 0.2$ & 5$^{+3}_{-2}$ & 0.8$^{+1.2}_{-0.5}$ \\
2A  &   & 362 & 3.3 & $+0.4$ & Tu-Elf-Owl & 
 &  &   & &  \\
2354$+$02 & 347$^{+13}_{-11}$ &  &   &  &  & 350 $\pm 25$ & 4.0 $\pm 0.25$ & 0.0 $\pm 0.2$ & 5$^{+3}_{-2}$ & 0.8$^{+1.2}_{-0.5}$ \\
2B  &   &  &  &  &  & 
 &  &   & &  \\
\enddata
\tablenotetext{a}{\citet{Beiler_2024} luminosity-based $T_{\rm eff}$.}
\tablenotetext{b}{``K-retrieval'', ``K-ATMO2020++'',``K-L-non-eq.-clear'', and ``K-Elf-Owl'' are \citet{Kuhnle_2024} retrieval, ATMO 2020++ grid \citep{Leggett_2021, Meisner_2023, Leggett_2025}, \citet{Lacy_2023} non-equilibrium clear atmosphere grid, and Elf Owl grid \citep{Mukherjee_2022} results, respectively. ``Tu-ATMO2020++'' and ``Tu-Elf-Owl'' are \citet{Tu_2024} ATMO 2020++ grid, and Elf Owl grid results, respectively.}
\tablenotetext{c}{The uncertainty in the atmospheric parameters is determined by the spacing of the ATMO 2020++ model grid.} 
\tablenotetext{d}{Mass and age are derived from the solar-metallicity evolutionary models of \citet{Marley_2021} adopting the $T_{\rm eff}$ and log $g$ values estimated here. The uncertainties in these two parameters are correlated in that a higher mass implies an older age. We have neglected the $\lesssim 10$\%  increase in age associated with enriched metallicity for 1047$+$54 and 1446$-$23 \citep{Marley_2021}. See also Figure 3.}
\end{deluxetable*}

\begin{figure}
\vskip -0.5in
\hskip 0.2in
\includegraphics[angle=0, width = 6 in]
{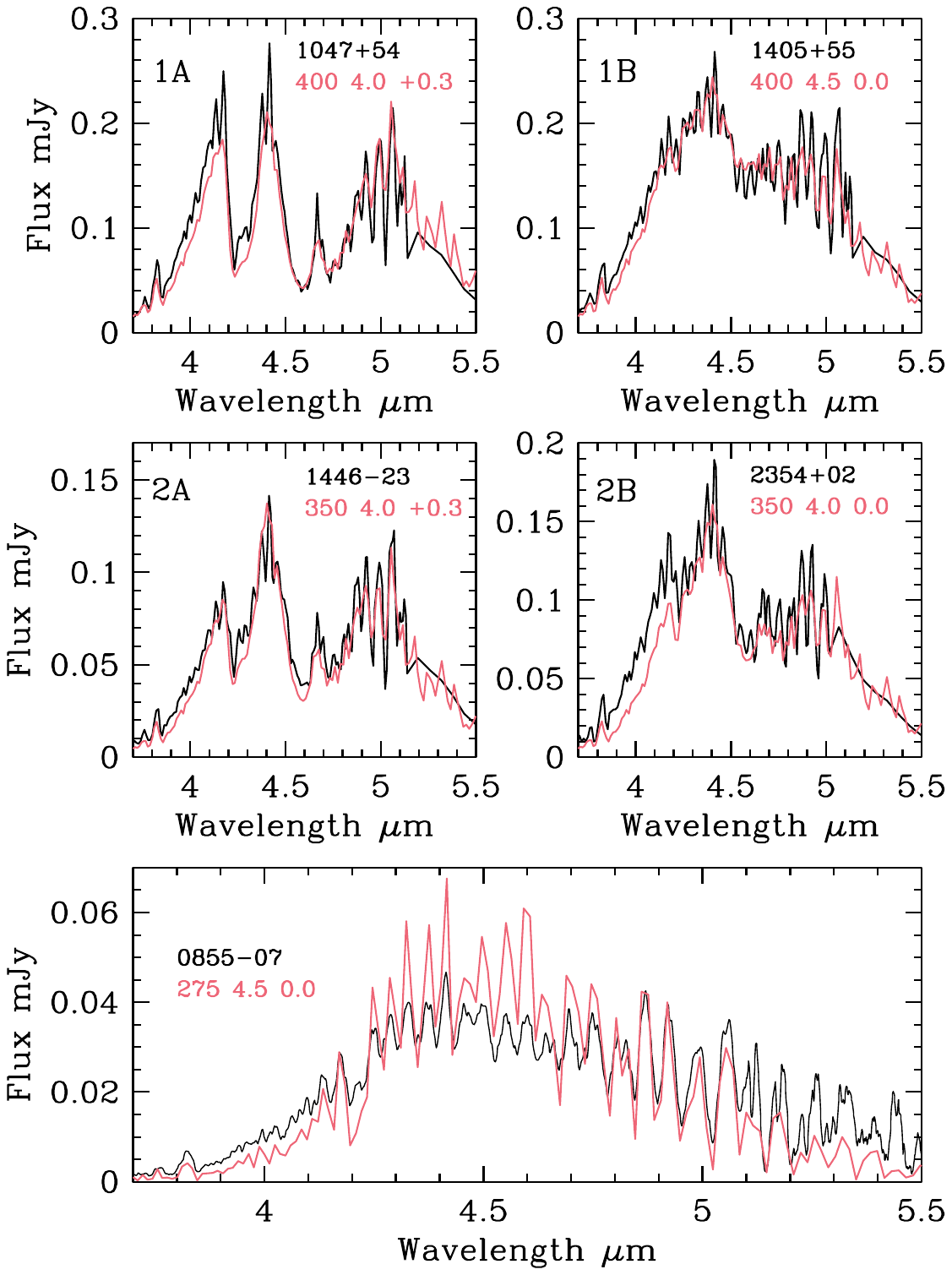}
\vskip -0.5in
\caption{Black lines are {\it JWST} spectra from \citet{Beiler_2024} for two 400~K Y dwarfs (1047$+$54 and 1405$+$55) and two 350~K Y dwarfs (1446$-$23 and 2354$+$02). Also shown in black is the higher resolution {\it JWST} spectrum for the 275~K Y dwarf 0855$-$07, from \citet{Kuhnle_2024}. Red lines are synthetic spectra from phosphine-free ATMO 2020++ models, with $T_{\rm eff}$, log $g$, and [m/H] shown in the legends. The synthetic spectra have been scaled by the measured distance to each source and the radius inferred from the temperature and gravity using \citet{Marley_2021} evolutionary models.   For comparison purposes the spectra are shown with all sources at 10~pc. 
}
\end{figure}

We find that the $4~\mu$m to $5~\mu$m spectral region provides key diagnostics for cold brown dwarf atmospheres. Figure 5 compares the observed spectra in that wavelength region for the five dwarfs, compared to the best fitting ATMO 2020++ synthetic spectra. 

The synthetic spectra were taken from the 
 \href{https://www.erc-atmo.eu/}{phosphine-free grid on the Opendata site}. The grid adopts an effective adiabat index $\gamma = 1.25$, and covers $250 \leq T_{\rm eff}$~K $\leq 1200$ (250~K, 275~K, 300~K, 350~K, 400~K, 450~K, 500K, then every 100~K to 1200~K), 
for three metallicities:  [m/H] $= -0.5, 0, +0.3$. The surface gravity $g$ ranges from log $g = 2.5$ to 
log $g = 5.5$ in steps of 0.5 dex. 
Out-of-equilibrium chemistry is used, with $K_{zz} = 10^5$ cm$^2$ s$^{-1}$ at log ~$g = 5.0$ and scaling by
$\times 10^{2(5 - {\rm log} ~g)}$ at other surface gravities to reflect scale height changes with gravity \citep[e.g.][]{Marley_2015}. 
More information on the models is given in the README file on the web site, and in \cite{Leggett_2025}.

The best-fit spectra have been selected by eye, given the coarseness of the grid.  Although the synthetic spectra systematically underestimate the flux  at wavelengths of 3 to 4$~\mu$m \citep[see discussion in][]{Leggett_2021}, the agreement is otherwise good.  Decreasing gravity or increasing metallicity significantly increases the absorption by CO$_2$ and CO at $4.2~\mu$m and $4.6~\mu$m respectively. Note that the degree of CO/CH$_4$ mixing is also impacted by the value of $K_{zz}$, which is  gravity-dependent. However \citet[][their Figure 4]{Zahnle_2014} show that for a 500~K dwarf, the increase in CO arising from an increase in $K_{zz}$ of a factor of 100
is insignificant compared to the abundance change brought about by a decrease in $g$  from $10^5$ to $10^4$ or an increase in [m/H] of 0.5 dex.

Our estimate of the properties of each dwarf, based on absolute brightness at $4.6~\mu$m and the strength of the absorption features between four and five microns, are consistent with other estimates determined either by grid fitting or retrieval techniques (Table 2).  The atmospheric temperature and gravity translate into masses and ages via evolutionary models, and this work produces ages which are consistent with the local brown dwarf population \citep[e.g.][]{Best_2024}. We note however that different models and different techniques can produce very different results (Table 2), showing that key pieces of our understanding of these cold and dynamic atmospheres are missing. We also point the reader to \citet{Wogan_2025} who describe an error in the calculation of the CO$_2$ abundance in the Elf Owl models. This will have impacted the estimates of gravity and metallicity in Table 2 which used the Elf Owl models.


\begin{figure}
\vskip -0.5in
\hskip 0.2in
\includegraphics[angle=0, width = 6 in]
{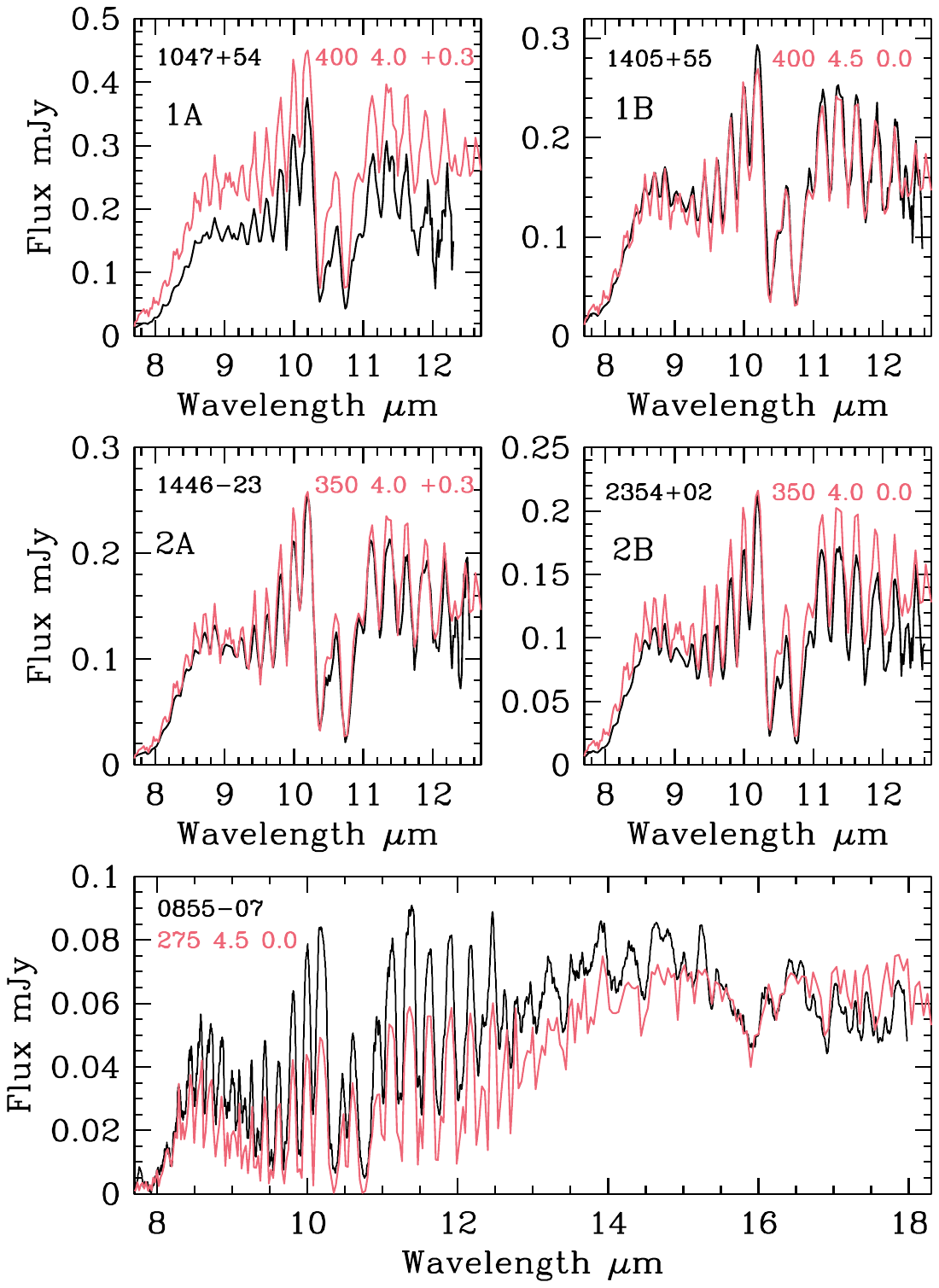}
\vskip -0.6in
\caption{Same as Figure 5 but for the $10~\mu$m wavelength region. 
}
\end{figure}

Figure 6 shows the {\it JWST} spectra for the same objects, and the same synthetic spectra, at longer wavelengths. Agreement is good for 1405$+$55, 1446$-$23, and 2354$+$02, but not as good for the low-gravity metal-rich Y dwarf 1047$+$54, or the cold Y dwarf 0855$-$07.  For 0855$-$07 it seems that specific changes to the atmospheric chemistry are required in order to reduce the NH$_3$ absorption at $9 \lesssim \lambda~\mu$m  $\lesssim 15$; 
further exploration is beyond the scope of the current paper.  For 1047$+$54 it appears that the value of the pressure-temperature profile adiabat needs to be adjusted so that the upper atmosphere (where the longer wavelength flux originates) is colder, and the deeper atmosphere (where the near-infrared flux originates) is warmer (Figures 6 and 7).

\begin{figure}
\vskip -0.5in
\hskip 0.2in
\includegraphics[angle=0, width = 6 in]
{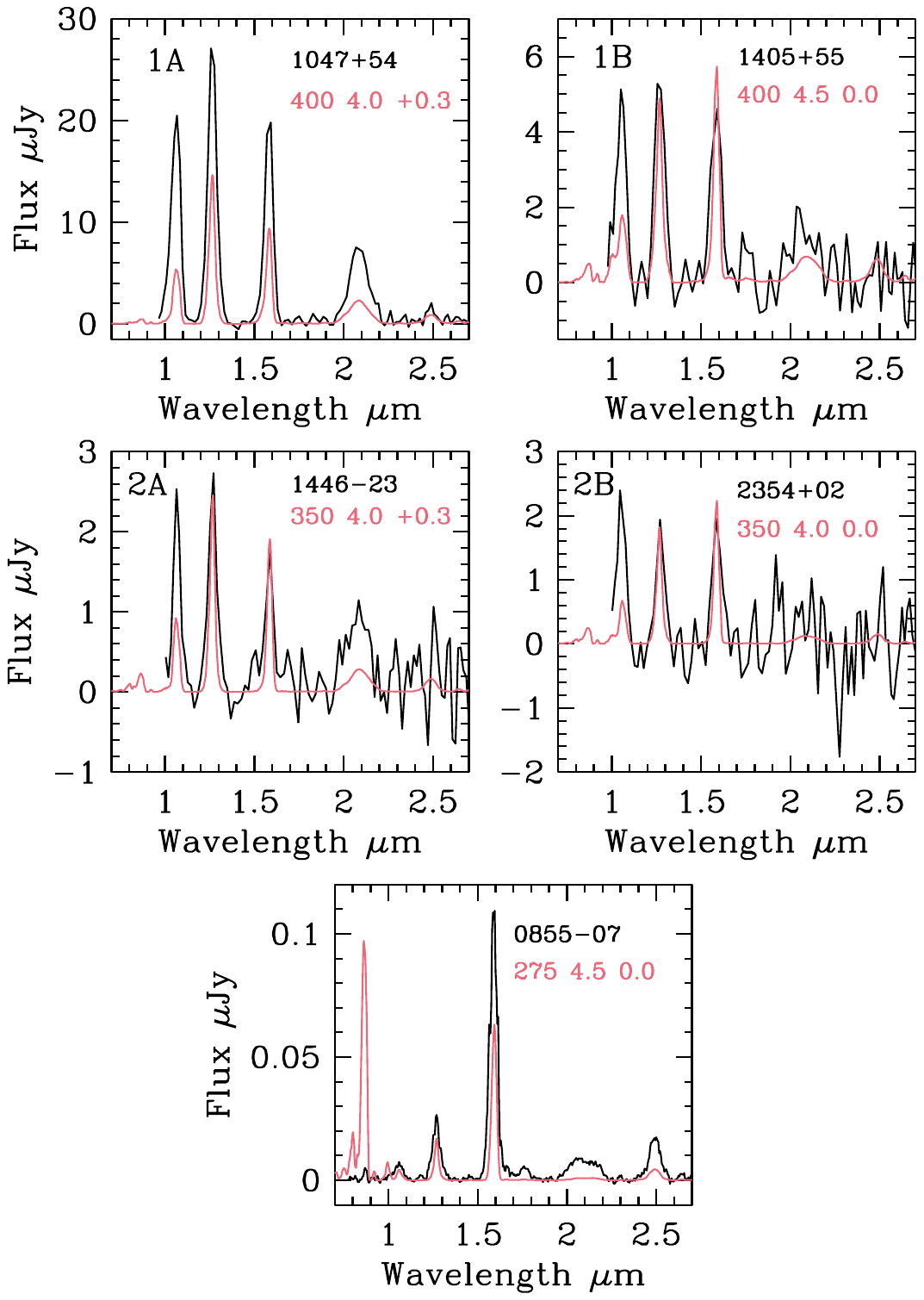}
\vskip -0.6in
\caption{Same as Figure 5 but for the near-infrared wavelength region. }
\end{figure}

Figure 7 shows the same spectral comparisons at near-infrared wavelengths, and Figure 8 shows a selection of near-infrared color diagrams. The models systematically under estimate the $Y$-band  ($\lambda \approx 1~\mu$m) flux. We found this in the fits to a larger sample of T and Y dwarfs in \citet{Leggett_2021} also, and we have attributed it to the uncertainty in the handling of the red wing of the resonance K~I line or an error in the abundance  of K~I due to sedimentation of alkali elements in the form of grains \citep{Phillips_2020}. The overly-faint $Y$ leads to the too-red $Y - J$ model colors seen in Figure 8.  The two 350~K  brown dwarfs, which differ in metallicity only according to the mid-infrared spectral fits, have significantly different $Y - J$ colors, more so than calculated by the models. The models do suggest that $Y - J$ becomes more sensitive to metallicity and gravity at cold temperatures, which appears to be supported by the data.  

\begin{figure}
\vskip -0.4in 
\hskip 0.2in
\includegraphics[angle=0, width = 6 in]
{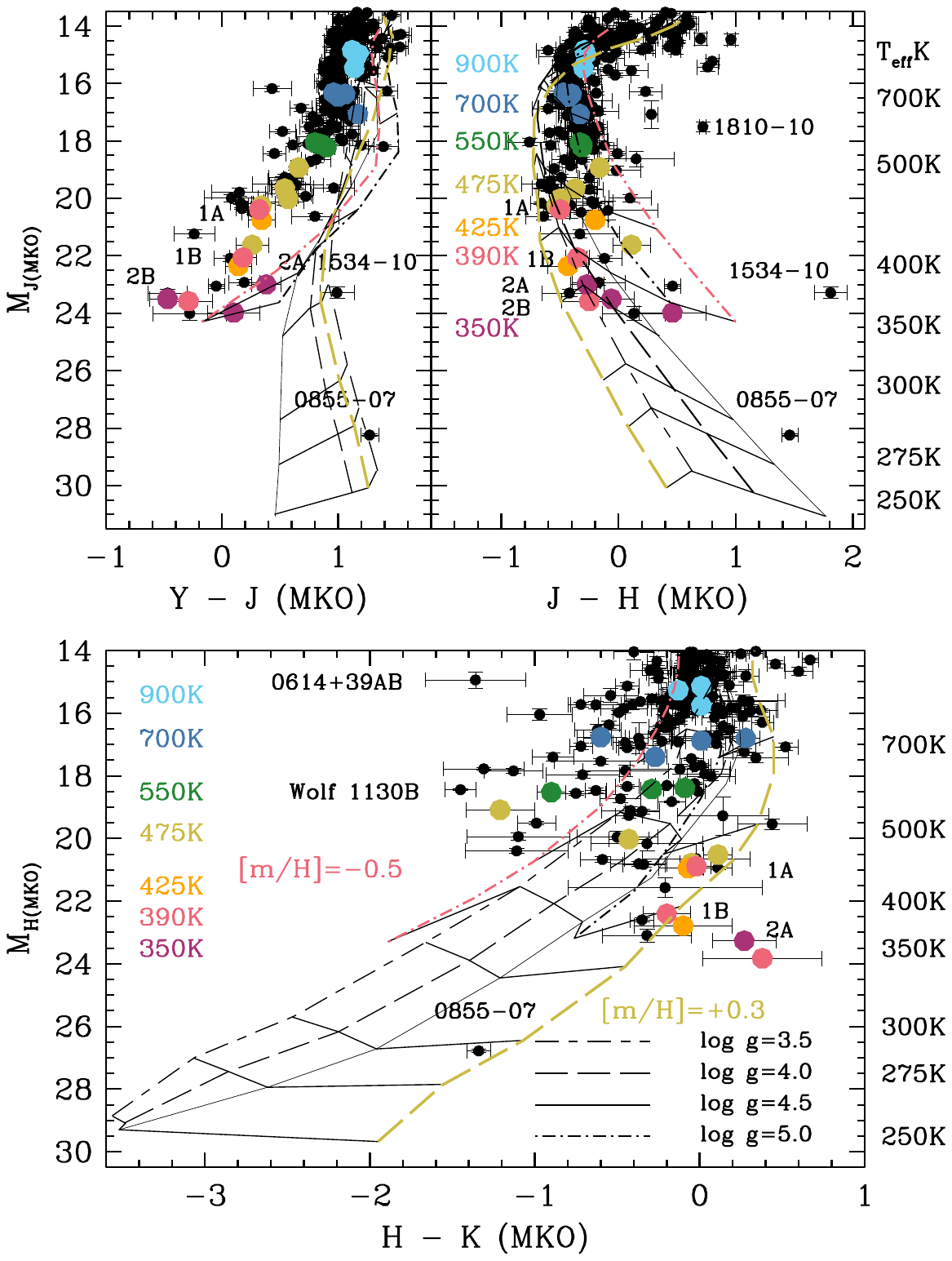}
\vskip -0.5in
\caption{Near-infrared color diagrams for the sample shown in Figure 2. ATMO 2020++ model sequences and isotherms are shown. \citet{Beiler_2024} $T_{\rm eff}$ values are indicated by colored symbols.  The colors of the two 400~K and the two 350~K objects are indicated as 1A and B, and 2A and B (Table 2).  These other objects are identified: the binary WISE 061407.49$+$391236.4AB  \citep{Kirkpatrick_2019};
the cold 0855$-$07 \citep{Luhman_2013}; and the metal-poor dwarfs 1534$-$10, 1810$-$10, and Wolf 1130B \citep{Kirkpatrick_2021b, Schneider_2020, Mace_2013b}. 
}
\end{figure}

In \citet{Leggett_2021} we also found that the models under estimated the flux at wavelengths between 2~$\mu$m and 4~$\mu$m, for colder Y dwarfs, and we find that here as well.  The flux at these wavelengths originates high in the atmosphere, and the discrepancy suggests that the atmospheres are warmer in the upper layers than calculated by the standard modified-adiabat ATMO 2020++ models. This discrepancy gives rise to the too-blue $H - K$ model colors for the coldest objects in Figure 8.

Figure 8 shows that the models calculate $J -H$ becoming redder with increasing gravity and with decreasing metallicity (see also the lower panel of Figure 2). However $H - K$, while also becoming redder with increasing gravity, becomes bluer with decreasing metallicity;  moreover the models show a larger dependence on metallicity than gravity for $H - K$. Trends in $J -H$ and $H - K$ therefore offer the possibility of disentangling metallicity and gravity effects. For example, Wolf 1130B stands out as blue in the $M_H$:$H - K$ diagram, but not in the $M_J$:$J - H$ diagram ($J - H = -0.01$),  while WISE 1810$-$10 stands out in $J - H$ but not $H - K$ ($H - K = -0.62$.)  The metallicity of WISE 1810$-$10 is
$\approx -1.5$ dex  \citep{Zhang_2025_1810}, while that of Wolf 1130B (based on its subdwarf primary) is 
$\approx -0.75$ dex  \citep{Kessel_2019}; if WISE 1810$-$10 has a higher gravity than Wolf 1130B that could redden the otherwise blue $H - K$; 
\citet{Zhang_2025_1810} estimate log $g = 5.5$ for WISE 1810$-$10 while \citet{Mace_2013a} estimate a range in log $g$ of 4.5 -- 5.5 for Wolf 1130B.

For the very cold 0855$-$07, the spike in flux in the best-fit model at $\lambda \approx 0.85~\mu$m seen in Figure 7 is not physical. It may be an artefact of the treatment of the opacities at these short wavelengths. 
Currently the H$_2$ opacity in the models does not extend to wavelengths shorter than $1.0~\mu$m, and the CH$_4$ and NH$_3$ opacity stops at  $0.8~\mu$m. The issue may also or instead be associated with the treatment of the strong K~I resonance doublet at $\lambda \approx 768$~nm. This deficiency is likely impacting the model flux distribution across the near-infrared region at very cold temperatures, as the agreement at $J$ and $H$ for 0855$-$07 is not as good as it is for the warmer Y dwarfs, and the model color extensions to cold temperatures in Figure 2 (lower panel) and Figure 8 are somewhat inconsistent with the adopted parameters for 0855$-$07 (275~K, log $g = 4.5$, [m/H] $= 0$)  which are based on a spectral fit (Figure 5).

Note that the models show a larger range in absolute $J$ and $H$ for given $T_{\rm eff}$ than in $M_{\rm W2}$ -- around $^+_- 1$ magnitude as opposed to $^+_- 0.2$ magnitude. The absolute near-infrared brightness is not a good indicator of temperature, while the absolute 4.6~$\mu$m brightness is (Figures 2 and 4).

\bigskip
\section{Euclid Photometry}

Table 3 gives new synthesized {\it Euclid} NISP photometry for late-type dwarfs, using the spectra referenced in the Table, on the AB system. Figures 9 and 10 are color-magnitude and color-color diagrams using the NISP magnitudes in Table 3, combined with those of \cite{Sanghi_2024}. ATMO 2020++  {\it Euclid} NISP magnitudes are given in the Appendix for $T_{\rm eff} =$ 275~K, 300~K, 350~K, 400~K, 450~K, and 500~K; log $g =$ 3.5. 4.0, 4.5, and 5.0; and [m/H] $=$ 0.0, $+0.3$, and $-0.5$.
The broader and redder bandpasses of the {\it Euclid} $YJH$ filters results in somewhat different color trends --- compare Figure 9 to Figure 8 (note however that the MKO system uses Vega magnitudes while {\it Euclid} uses AB). The unusual near-infrared spectrum of the very metal-poor 1534$-$10 (Faherty et al. 2025, Nature in press) produces a much brighter $J_E$ and fainter $H_E$ compared to the MKO filters, so that this object does not appear red in $J_E - H_E$ but is very red in the equivalent MKO and Roman colors.

\startlongtable
\begin{deluxetable*}{lcrrrc}
\tablecaption{Synthesized {\it Euclid} Near-Infrared Photometry, AB mags}
\tablehead{
\colhead{AllWISE Name} & 
\colhead{Spec.} &
\colhead{$Y_E$} & 
\colhead{$J_E$} &
\colhead{$H_E$} & 
\colhead{Spectral} \\
\colhead{} & 
\colhead{Type} & 
\colhead{} & 
\colhead{} &
\colhead{} & 
\colhead{Source} 
}
\startdata
014807.25$-$720258.7	&	T9.5		&	21.01	$\pm$	0.05	&	20.57	$\pm$	0.03	&	21.15	$\pm$	0.05	&	Cu11 	\\
024714.52$+$372523.5	&	T8	&		20.20	$\pm$	 0.05 	&	19.54	$\pm$	 0.05 	&	20.16	$\pm$	 0.05 	&	Be24	\\
031325.96$+$780744.2	&	T8.5		&	19.47	$\pm$	0.05 	&	19.01	$\pm$	 0.05 	&	19.57	$\pm$	 0.05 	&	Be24	\\
032504.33$-$504400.3 	&	T8	&		21.10	$\pm$	0.05	&	20.49	$\pm$	0.03	&		& Sc15	\\
033515.01$+$431045.1	&	T9		&	21.46	$\pm$	0.10	&	21.07	$\pm$	0.03	&		&	Sc15	\\
034807.72$-$602227.0 &	T7		& 17.02 $\pm$  0.05 & 16.53  $\pm$ 0.05  & 17.11 $\pm$ 0.05 & R-GTO	\\
035934.06$-$540154.6	&	Y0		&	23.11	$\pm$	 0.08 	&	22.99	$\pm$	 0.06 	&	23.44	$\pm$	 0.06 	&	Be23	\\
040443.48$-$642029.9	&	T9	&		21.60	$\pm$	0.06	&	21.19	$\pm$	0.02	&		&	Sc15	\\
041022.71$+$150248.5	&	Y0	&		21.20	$\pm$	0.06	&	21.12	$\pm$	0.05	&	21.70	$\pm$	0.07	&	Le17	\\
041358.14$-$475039.3	&	T9	&		21.69	$\pm$	0.15	&	21.08	$\pm$	0.06	&	21.60	$\pm$	0.10	&	Ma13	\\
043052.92$+$463331.6	&	T8	&		20.86	$\pm$	 0.05 	&	20.54	$\pm$	 0.05 	&	21.04	$\pm$	 0.05 	&	Be24	\\
053516.80$-$750024.9	&	Y1	&		24.10	$\pm$	0.20	&	23.96	$\pm$	 0.13 	&	24.03	$\pm$	 0.16 	&	Be24	\\
064723.23$-$623235.5 	&	Y1	&		24.54	$\pm$	0.25	&	24.72	$\pm$	0.08	&		& Sc15	\\
073444.02$-$715744.0 	&	Y0	&		22.13	$\pm$	 0.05 	&	21.89	$\pm$	 0.05 	&	22.51	$\pm$	 0.05 	&	Be24	\\
075108.79 -763449.6$^*$	&	T9	&	 21.18	$\pm$	 0.07	&	 20.85	$\pm$	 0.05	&	 21.14	$\pm$	0.08	&  Ki11, Zh25	\\	
082507.35$+$280548.5	&	Y0.5		&	23.88	$\pm$	0.15	&	24.22	$\pm$	0.11	&	24.50	$\pm$	0.25	&	Be24, Sc15	\\
085510.83$-$071442.5	&	Y4	&		27.59	$\pm$	 0.05 	&	26.31	$\pm$	 0.05 	&	25.41	$\pm$	 0.05 	&	Lu24 \\
085834.42$+$325627.7	&	T1	&		17.96	$\pm$	0.03	&	17.46	$\pm$	0.02	&	17.00	$\pm$	0.02	&	Bg10	\\
104756.81$+$545741.6	&	Y0	&		22.86	$\pm$	 0.06 	&	22.88	$\pm$	 0.05 	&	23.53	$\pm$	 0.05 	&	Be24	\\
114156.67$-$332635.5 	&	Y0	&	21.74	$\pm$	0.15	&	21.25	$\pm$	0.06	&	&	Le17	\\
120604.38$+$840110.6	&	Y0	&		22.21	$\pm$	 0.06 	&	22.00	$\pm$	 0.05 	&	22.58	$\pm$	 0.05 	&	Be24	\\
121756.91$+$162640.2A	&	T9	&		19.88	$\pm$	0.07	&	19.50	$\pm$	0.03	&	20.13	$\pm$	0.03	&	Le16	\\
121756.91$+$162640.2B	&	Y0	&		21.61	$\pm$	0.14	&	21.85	$\pm$	0.09	&	22.04	$\pm$	0.09	&	Le16	\\
133553.45$+$113005.2 	&	T8.5	&		&	19.37	$\pm$	0.02	&	20.03	$\pm$	0.04	&	Bn08	\\
140518.40$+$553421.4	&	Y0.5	&	22.78	$\pm$	0.08	&	22.73	$\pm$	0.08	&	23.01	$\pm$	 0.13 	&	Be24, Sc15	\\
141623.94$+$134836.3	&	T7.5		&	&	18.82	$\pm$	0.02	&	19.48	$\pm$	0.02 	&	Bn10	\\
144606.62$-$231717.8 	&	Y1	&		24.59	$\pm$	 0.12	&	24.57	$\pm$	 0.07 	&	24.99	$\pm$	 0.16 	&	Be24	\\
150115.92$-$400418.4	&	T6	&		18.14	$\pm$	 0.05 	&	17.67	$\pm$	 0.05 	&	18.18	$\pm$	 0.05 	&	Be24	\\
153429.75$-$104303.3	&	T5	&		26.24	$\pm$	 0.07 	&	24.20	$\pm$	 0.13 	&	24.62	$\pm$	 0.05 	&	Fa25	\\
154151.66$-$225025.2	&	Y1	&		22.69	$\pm$	 0.09 	&	22.89	$\pm$	 0.11 	&	23.45	$\pm$	0.20	&	Be24	\\
154214.00$+$223005.2	&	T9.5		&	21.79	$\pm$	0.05	&	21.55	$\pm$	0.03	&	&	Ma13	\\
163940.86$-$684744.6	&	Y0	&		22.03	$\pm$	0.04	&	21.99	$\pm$	0.03	&	&	Sc15	\\
173835.53$+$273258.9 	&	Y0	&		&	21.32	$\pm$	0.12	&	21.94	$\pm$	0.15	&	Cu11, Le16	\\
181006.00$-$101001.1	&	T8	&		20.08	$\pm$	0.05	&	19.50	$\pm$	0.02	&	19.46	$\pm$	0.02	&	Sc20, Lo22	\\
182831.08$+$265037.8	&	Y2	&		24.42	$\pm$	0.05 	&	24.52	$\pm$ 0.05 	&	24.37	$\pm$	 0.05	&	R-GTO	\\
200520.38$+$542433.9$^{**}$ 	&	T8	&	21.35	$\pm$	0.05	&	21.06	$\pm$	0.04	&	&	Ma13	\\
205628.90$+$145953.3	&	Y0	&		21.06	$\pm$	 0.05 	&	20.79	$\pm$	 0.05 	&	21.44	$\pm$	 0.05 	&	Be24	\\
210200.15$-$442919.5	&	T9	&		20.19	$\pm$	 0.05	&	19.76	$\pm$	 0.05 	&	20.36	$\pm$	 0.05 	&	Be24	\\
214638.83$-$001038.7	&	T8.5	&		&	19.79	$\pm$	0.03	&	20.48	$\pm$	0.03	&	Bn09	\\
215949.54$-$480855.2 	&	T9	&		20.89	$\pm$	 0.05	&	20.45	$\pm$	 0.05 	&	21.04	$\pm$	 0.05 	&	Be24	\\
221216.33$-$693121.6	&	T9	&		21.57	$\pm$	0.03	&	21.36	$\pm$	0.03	&	&	Sc15	\\
222055.31$-$362817.4	&	Y0	&		22.42	$\pm$	0.07	&	22.21	$\pm$	0.03	&	&	Sc15	\\
232519.54$-$410534.9	&	T9	&		21.39	$\pm$	0.12	&	21.14	$\pm$	0.03	&	21.65	$\pm$	0.10	&	Ma13	\\
235402.79$+$024014.1 	&	Y1	&		24.05	$\pm$	 0.16 	&	24.76	$\pm$	 0.18 	&	24.49	$\pm$	0.20	&	Be24	\\
\enddata
\tablenotetext{*} {Also known as COCONUT2-2b.}
\tablenotetext{** } {\ Also known as Wolf 1130B.}
\tablerefs{
Bg10 -- \citet{Burgasser_2010};
Bn08 -- \citet{Burningham_2008};
Bn09 -- \citet{Burningham_2009};
Bn19 -- \citet{Burningham_2010a};
Be23 -- \citet{Beiler_2023}; 
Be24 -- \citet{Beiler_2024}; 
Cu11 -- \citet{Cushing_2011}; 
Cu21 -- \citet{Cushing_2021};
Fa25 -- Faherty et al. 2025, Nature in press;
Ki11 -- \citet{Kirkpatrick_2011};
Le16 -- \citet{Leggett_2014};
Le16 -- \citet{Leggett_2016a}; 
Le17 -- \citet{Leggett_2017};
Lo22 -- \citet{Lodieu_2022}; 
Lu24 -- \citet{Luhman_2024};
Ma13 -- \citet{Mace_2013a}; 
R-GTO -- GTO 1189, PI = Roellig; 
Sc15 -- \citet{Schneider_2015};
Sc20 -- \citet{Schneider_2020};
Zh25 -- \citet{Zhang_2025_COCO}.
}
\end{deluxetable*}

\begin{figure}
\vskip -0.1in
\hskip 0.15in
\includegraphics[angle=-90, width = 7.5 in]{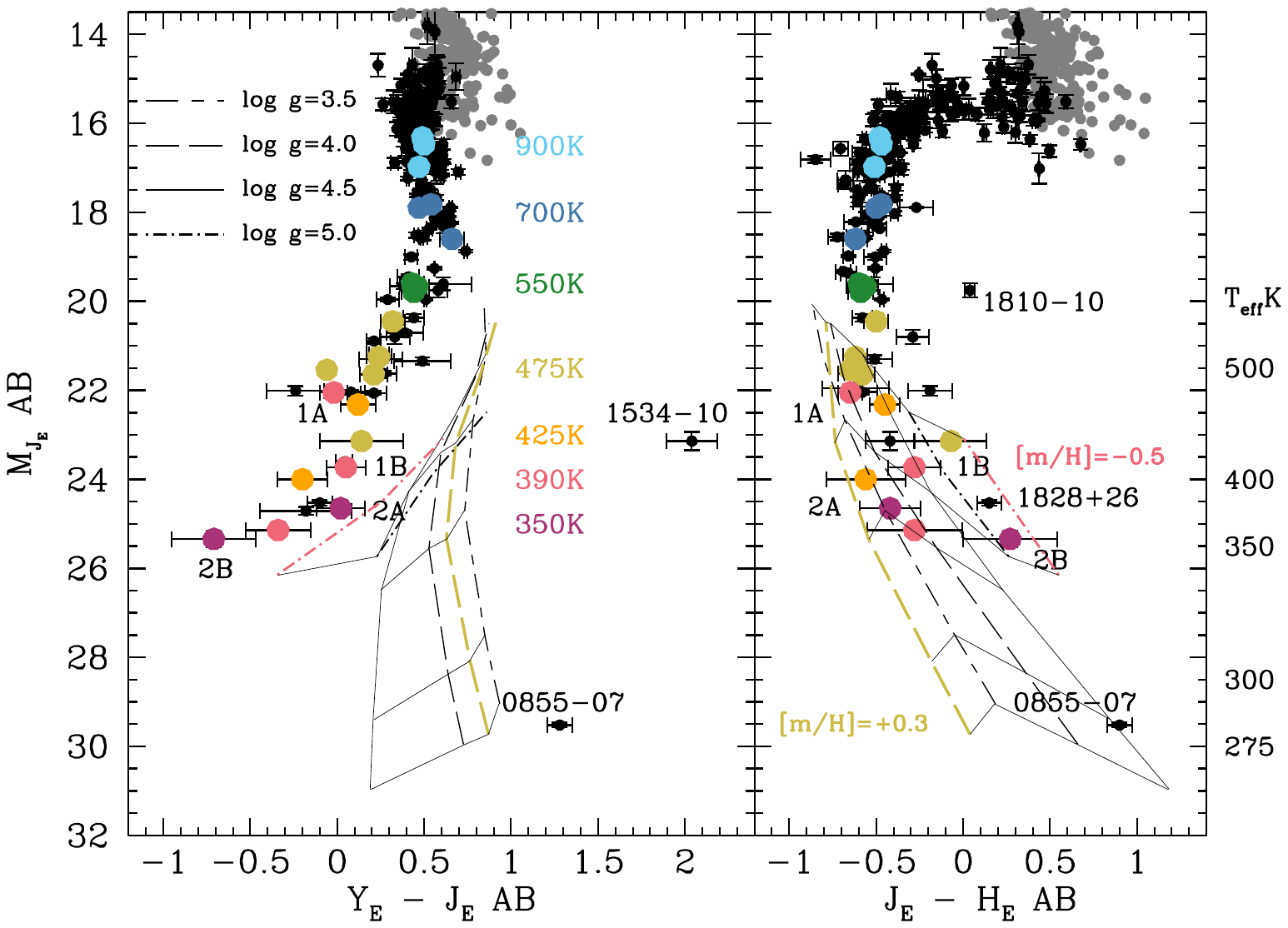}
\vskip -0.1in
\caption{Symbols are {\it Euclid} colors from \citet{Sanghi_2024} and Table 3. Grey dots are M and L dwarfs, black dots are T and Y dwarfs. Colored symbols are dwarfs in the \citet{Beiler_2024} sample, where the color indicates the Beiler et al. $T_{\rm eff}$ value, shown in the legend in the left panel.  Red, yellow, and black almost vertical lines are ATMO 2020++ sequences with model parameters shown in the legends. Lines across the sequences are isotherms with  $T_{\rm eff}$ values indicated on the right axis. The systematic offset in $Y_E - J_E$ is due to the model spectra being too faint at $Y$, see Section 4 and Figure 7.
The 275~K Y dwarf 0855$-$07 is identified, as are the metal-poor dwarfs 1534$-$10 and 1810$-$10. 1828$+$26 is likely to be a high-gravity unresolved binary \citep{Leggett_2025}.  
}
\end{figure}

Figure 10 shows that it will be difficult to identify cold Y dwarfs in {\it Euclid} by NISP colors alone, that is, without a luminosity estimate. 
However identifying point sources with proper motion will be one way to identify nearby sources.
The {\it Euclid} Wide Field Survey has an estimated depth in the near-infrared of 24.5$_{AB}$ \citep{Euclid_WideSurvey}, meaning that any object colder than 400~K will need to be closer than 10~pc (Figure 9). \citet{Euclid_UCD} use {\it Euclid} spectra to identify cool dwarfs in the Deep Field North. They show that in color-space the primary point-source contaminant is high redshift QSOs. The cool dwarfs however will have a proper motion, and will be redder in VIS - near-infrared colors \citep[][their Figures 7 and 8]{Euclid_UCD}.

\begin{figure}
\vskip -0.3in
\hskip 0.3in
\includegraphics[angle=-90, width = 6 in]{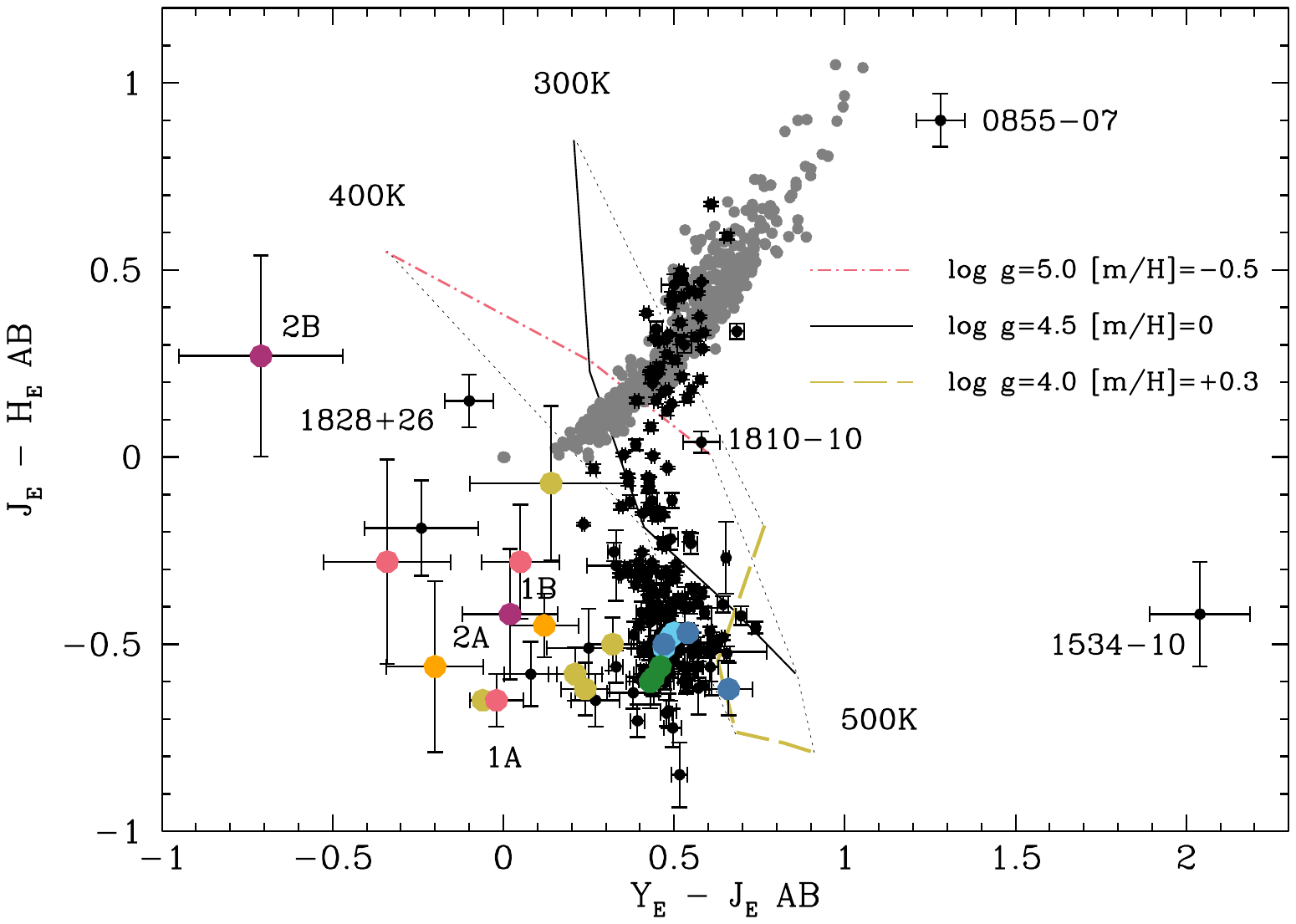}
\vskip -0.1in
\caption{{\it Euclid} color-color diagram with symbols as in Figure 9. Three model sequences are shown which sample the range of model colors in Figure 9. Parameters are given in the legend. The log $g = 4.0$ and 4.5 sequences span $300 \leq T_{\rm eff} \leq 500$, and the 
log $g = 5.0$ sequence spans $400 \leq T_{\rm eff} \leq 500$ (colder high gravity dwarfs are older than 15~Gyr \citep[and Figure 3]{Marley_2021}). Dotted lines join the 300, 400, and 500~K points across model sequences. The models are flux-deficient in the $Y_E$ band, resulting in a red offset in $Y_E - J_E$. 1828$+$26 is likely to be a high-gravity  unresolved binary \citep{Leggett_2025}. 1534$-$10 and 1810$-$10 are extremely metal-poor with 
[m/H] $\approx -2.2$ and $-1.5$ dex respectively 
\citep[Faherty et al. 2025, Nature in press and][]{Zhang_2025_1810}.
}
\end{figure}

\begin{deluxetable*}{lcrcrr}
\tablecaption{Synthesized {\it Euclid} VIS $I_E$ Photometry, AB mags}
\tablehead{
\colhead{AllWISE Name} & 
\colhead{Spectral} &
\colhead{$I_E$} & 
\colhead{Spectral} & \colhead{$I_E - Y_E$} & \colhead{$I_E - H_E$} \\
\colhead{} & 
\colhead{Type} &
\colhead{} & 
\colhead{Source} &  & 
}
\startdata
015141.69+124429.6 & T1 &  19.97 $\pm$ 0.07 & Bg04 & 2.01 $\pm$ 0.07 & 2.83  $\pm$ 0.07 \\
041519.50$-$093506.0 & T8 & 22.01 $\pm$ 0.03 & Bg03 & 4.63 $\pm$ 0.03	& 4.60 $\pm$ 0.03 \\
072227.51$-$054031.2 & T9 & 23.04 $\pm$ 0.10  & Le12 & 4.58 $\pm$ 0.10	& 4.39 $\pm$ 0.10\\
072718.20$+$171001.0 & T7 & 21.74 $\pm$ 0.03 & Bg03 & 4.59 $\pm$ 0.03 	& 4.35 $\pm$ 0.03 \\
085510.83$-$071442.5 & Y4 &   29.56 $\pm$ 0.25 & Lu24 &  1.97 $\pm$  0.25	&  4.15 $\pm$ 0.25\\
120956.33$-$100405.0 	& T3 	& 20.84 $\pm$  0.20 & Bg04 & 3.59  $\pm$  0.20 & 4.05  $\pm$  0.20 \\
125453.90$-$012247.4 & T2 & 20.36 $\pm$ 0.02 & Le00 & 3.91 $\pm$ 0.02 & 4.69 $\pm$ 0.02 \\
134646.45$-$003150.4 & T6.5 & 21.68 $\pm$ 0.03 & Bg00  & 4.25  $\pm$ 0.03 & 	5.27 $\pm$ 0.03 \\
145714.96$-$212147.7 & T7.5 & 21.66 $\pm$ 0.03 & Bg03 & 4.97  $\pm$ 0.03 	& 4.68  $\pm$ 0.03 \\
150319.68+252525.7 & T5 & 20.44 $\pm$ 0.10 & Bg03 & 4.88  $\pm$ 0.10 & 4.88  $\pm$ 0.10 \\
162414.37$+$002915.6 & T6 & 21.58 $\pm$ 0.03  & Bg00 & 4.43 $\pm$ 0.03 	& 4.45 $\pm$ 0.03 \\
175023.85$+$422237.3 & T1 & 19.90 $\pm$ 0.11 & Bg06 & 2.05 $\pm$ 0.11 & 	2.81 $\pm$ 0.11 \\
205628.90$+$145953.3 & Y0	& 25.02 $\pm$ 0.07  & Le13 & 4.28 $\pm$  0.09	& 3.57 $\pm$  0.09 \\
214428.47$+$144607.8 & T2.5 & 19.87 $\pm$ 0.05  & Lu07 & 2.28  $\pm$ 0.05 & 	2.93 $\pm$ 0.05 \\
225418.96+312352.2 & T4 & 21.22 $\pm$ 0.15 & Bg04, Kn04 & 4.31 $\pm$ 0.15 & 4.76 $\pm$ 0.15 \\
\enddata
\tablerefs{
Bg00 -- \citet{Burgasser_2000};
Bg03 -- \citet{Burgasser_2003}; 
Bg04 -- \citet{Burgasser_2004};
Bg06 -- \citet{Burgasser_2006a};
Kn04 -- \citet{Knapp_2004};
Le00 -- \citet{Leggett_2000}
Le12 -- \citet{Leggett_2012}; 
Le13 -- \citet{Leggett_2013};
Lu07 -- \citet{Luhman_2007};
Lu24 -- \citet{Luhman_2024}.}
\end{deluxetable*}

\begin{figure}
\vskip -1.0in
\hskip 1.1in
\includegraphics[angle=0, width = 4.2 in]{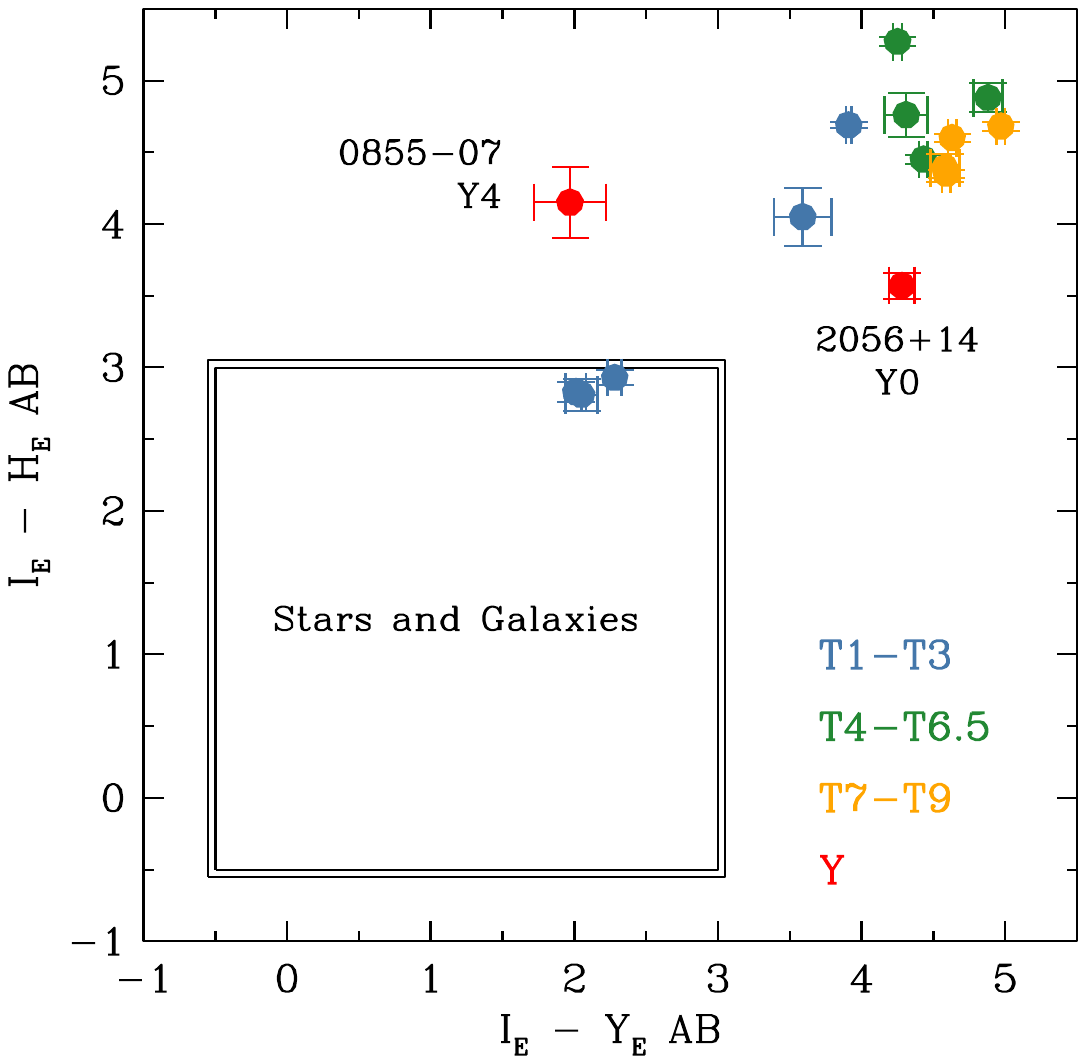}
\vskip -0.5in
\caption{Schematic {\it Euclid} optical to near-infrared color-color diagram. The box illustrates  the location of stars and galaxies as shown by \citet{Euclid_AGN, Euclid_highZ, Euclid_UCD}. Dots are T and Y dwarfs from Table 4 where color indicates spectral type. The Y dwarfs are identified by shortened RA-Decl.. }
\end{figure}

There are not many late-T and Y dwarfs with red optical spectra; we used spectra of T dwarf near-infrared spectral standards from \citet{Burgasser_2004,Burgasser_2006b}, and added later types where red spectra were available, to synthesize VIS $I_E$ photometry (Table 4). Figure 11 illustrates the location of stars, galaxies and T and Y dwarfs in the visible to near-infrared color space.  The $I - Y$ and $I - H$ colors of late-T and Y dwarfs are redder than those of AGN and spiral galaxies \citep[][their Figure 3]{Euclid_AGN}, and so this can be used to differentiate between the two populations, if a limit on $I_E$ can be obtained. Note that the $I - Y$ and $I - H$ colors of T dwarfs generally become redder with later spectral types, but this trend reverses for the colder Y dwarfs.

\clearpage
\section{Roman Photometry}

Table 5 gives new synthesized photometry in the {\it Roman} filters, using the spectral sources referenced in the Table, on the AB system. Figures 12 and 13 are color-magnitude and color-color diagrams using these data, together with those presented by \citet{Sanghi_2024}. ATMO 2020++  {\it Roman} magnitudes are given in the Appendix for $T_{\rm eff} =$ 275~K, 300~K, 350~K, 400~K, 450~K, and 500~K; log $g =$ 3.5. 4.0, 4.5, and 5.0; and [m/H] $=$ 0.0, $+0.3$, and $-0.5$.

\startlongtable
\begin{deluxetable*}{lcrrrrrc}
\tablecaption{Synthesized {\it Roman} Near-Infrared Photometry, AB mags}
\tablehead{
\colhead{AllWISE Name} & 
\colhead{Spec.} &
\colhead{F106} & 
\colhead{F129} &
\colhead{F158} & 
\colhead{F184} & 
\colhead{F213} & 
\colhead{Spectral} \\
\colhead{} & 
\colhead{Type} & 
\colhead{} & 
\colhead{} &
\colhead{} & 
\colhead{} & 
\colhead{} & 
\colhead{Source} 
}
\startdata
005910.90$-$011401.3	&	T8.5	&				&	19.35	$\pm$	0.03	&	19.81	$\pm$	0.03	&	21.11 	$\pm$ 0.30 & 20.17	$\pm$	0.10	&	De08\\	
014807.25$-$720258.7	&	T9.5	&	21.07	$\pm$	0.05	&	20.65	$\pm$	0.03	&	21.13	$\pm$	0.05	&	&  21.44	$\pm$	0.20	&	Cu11 	\\
024714.52$+$372523.5	&	T8.5	&	20.31	$\pm$	 0.05		&	19.62	$\pm$	 0.05		&	20.06	$\pm$	 0.05		&	21.44	$\pm$	 0.07	 & 20.37	$\pm$	 0.07		&	Be24	\\
030449.03$-$270508.3	&	Y0	&			&	24.04	$\pm$	0.08	&	24.50	$\pm$	0.12	&	&			&	Pi14	\\
031325.96$+$780744.2	&	T8.5	&	19.55	$\pm$	 0.05		&	19.07	$\pm$	 0.05		&	19.50	$\pm$	 0.05		& 21.00 $\pm$	 0.07	 & 19.82	$\pm$	 0.06		&	Be24	\\
032504.33$-$504400.3 	&	T8	&	21.18	$\pm$	0.05	&	20.56	$\pm$	0.03	&	21.06	$\pm$	0.05	&	& 			&	Sc15	\\
033515.01$+$431045.1	&	T9	&	21.53	$\pm$	0.10	&	21.14	$\pm$	0.03	&	21.55	$\pm$	0.05	&		& 		&	Sc15	\\
034807.72$-$602227.0 &	T7		& 17.10 $\pm$ 0.05 & 16.56  $\pm$ 0.05 & 17.06 $\pm$ 0.05 & 18.03 $\pm$ 0.05 & 17.24 $\pm$ 0.05	 & R-GTO	\\
035000.32$-$565830.2	&	Y1	&				&	23.78	$\pm$	0.09	&	23.38	$\pm$	0.14	&				& &	Le16	\\
035934.06$-$540154.6	&	Y0	&	23.17	$\pm$	 0.08		&	23.09	$\pm$	 0.06		&	23.37	$\pm$	 0.06	 & 	24.88 $\pm$ 0.30 &	23.52	$\pm$	0.25 &	Be23	\\
040443.48$-$642029.9	&	T9	&	21.64	$\pm$	0.06	&	21.31	$\pm$	0.02	&	21.43	$\pm$	0.03	&		& 		&	Sc15	\\
041022.71$+$150248.5	&	Y0	&	21.25	$\pm$	0.06	&	21.12	$\pm$	0.05	&	21.75	$\pm$	0.07	&	& 21.85	$\pm$	0.20	&	Le17	\\
041358.14$-$475039.3	&	T9	&	21.72	$\pm$	0.15	&	21.17	$\pm$	0.06	&	21.68	$\pm$	0.10	&	& 			&	Ma13	\\
043052.92$+$463331.6	&	T8	&	20.96	$\pm$	 0.05		&	20.61	$\pm$	 0.05		&	20.91	$\pm$	 0.05		& 22.29 $\pm$	 0.11	 & 	22.10	$\pm$	 0.11		&	Be24	\\
053516.80$-$750024.9	&	Y1	&	24.18	$\pm$	0.20	&	24.12	$\pm$	 0.13		&	24.15	$\pm$	 0.16		&	& 			&	Be24	\\
064723.23$-$623235.5 	&	Y1	&	24.55	$\pm$	0.25	&	24.87	$\pm$	0.08	&				&		& 		&	Sc15	\\
071322.55$-$291751.9	&	Y0	&				&	21.64	$\pm$	0.05	&	22.38	$\pm$	0.10	&			& 	&	Le17	\\
073444.02$-$715744.0 	&	Y0	&	22.20	$\pm$	 0.05		&	21.95	$\pm$	 0.05		&	22.43	$\pm$	 0.05	 & 24.00 $\pm$ 0.25	&	22.89	$\pm$	 0.11		&	Be24	\\
075108.79$-$763449.6$^*$	&	T9	&	 21.21	$\pm$	 0.07	&	 20.90	$\pm$ 	0.05	&	21.26	$\pm$	0.10	& &  21.05	$\pm$	0.25	& Ki11, Zh25	\\	
082507.35$+$280548.5	&	Y0.5	&	23.92	$\pm$	0.15	&	24.41	$\pm$	0.11	&	24.68	$\pm$	0.25	&	& 		&	Be24, Sc15	\\
085510.83$-$071442.5	&	Y4	&	27.58	$\pm$	 0.05	&	26.65	$\pm$	 0.05		&	25.32	$\pm$	 0.05		& 27.47	$\pm$	 0.08	 &	26.63	$\pm$	 0.06		&	Lu24	\\
085834.42$+$325627.7	&	T1	&	18.02	$\pm$	0.03	&	17.59	$\pm$	0.02	&	17.05	$\pm$	0.02	& 17.11 $\pm$	0.03 &	16.68	$\pm$	0.03	&	Bg10	\\
104756.81$+$545741.6	&	Y0	&	22.90	$\pm$ 0.06		&	22.87	$\pm$	 0.05		&	23.49	$\pm$	 0.05		& 25.49	$\pm$  0.11		&	23.54	$\pm$  0.07			&	Be24	\\
114156.67$-$332635.5 	&	Y0	&	21.80	$\pm$	0.15	&	21.30	$\pm$	0.06	&	21.91	$\pm$	0.15	&	& 			&	Le17	\\
120604.38$+$840110.6	&	Y0	&	22.25	$\pm$	 0.06		&	22.01	$\pm$	 0.05		&	22.62	$\pm$	 0.05	& 23.78 $\pm$	0.20 &	22.53	$\pm$	 0.11		&	Be24	\\
121756.91$+$162640.2A	&	T9	&	19.96	$\pm$	0.07	&	19.56	$\pm$	0.03	&	20.06	$\pm$	0.03	& 21.36 $\pm$ 0.15	 &	20.87	$\pm$	0.08	&	Le14	\\
121756.91$+$162640.2B	&	Y0	&	21.65	$\pm$	0.14	&	21.87	$\pm$	0.09	&	22.36	$\pm$	0.09	& &			&	Le14	\\
133553.45$+$113005.2 	&	T8.5	&				&	19.44	$\pm$	0.02	&	19.94	$\pm$	0.04	&	21.40 $\pm$	0.17 & 20.06	$\pm$	0.09	&	Bn08	\\
140518.40$+$553421.4	&	Y0.5	&	22.77	$\pm$	0.08	&	22.72	$\pm$	0.08	&	22.84	$\pm$	0.12	& &	23.13	$\pm$	0.30	&	Be24, Sc15	\\
141623.94$+$134836.3	&	T7.5	&				&	18.87	$\pm$	0.02	&	19.34	$\pm$	0.02	&	20.59 $\pm$	0.14 & 20.45	$\pm$	0.10	&	Bn10	\\
144606.62$-$231717.8 	&	Y1	&	24.63	$\pm$	 0.12		&	24.67	$\pm$	 0.07		&	24.95	$\pm$	 0.16	& 26.06	$\pm$	0.30 &	24.59	$\pm$	0.24	&	Be24	\\
150115.92$-$400418.4	&	T6	&	18.26	$\pm$	 0.05		&	17.70	$\pm$	 0.05		&	18.14	$\pm$	 0.05		& 18.89 $\pm$	 0.05	 &	18.21	$\pm$	 0.05	&	Be24	\\
153429.75$-$104303.3	&	T5	&	26.22	$\pm$	 0.07		&	24.91	$\pm$	 0.13		&	23.89	$\pm$	 0.05		& &				&	Fa25	\\
154151.66$-$225025.2	&	Y1	&	22.73	$\pm$	 0.09		&	22.87	$\pm$	 0.11		&	23.44	$\pm$	0.20	& &	23.35	$\pm$	0.30	&	Be24	\\
154214.00$+$223005.2	&	T9.5	&	21.84	$\pm$	0.05	&	21.59	$\pm$	0.03	& &				&		&	Ma13	\\
163940.86$-$684744.6	&	Y0	&	22.07	$\pm$	0.04	&	22.05	$\pm$	0.03	&	22.24	$\pm$	0.03	& &				&	Sc15	\\
173835.53$+$273258.9 	&	Y0	&			&	21.33	$\pm$	0.12	&	21.97	$\pm$	0.15	& &	21.39	$\pm$	0.10	&	Cu11, Le16	\\
181006.00$-$101001.1	&	T8	&	20.13	$\pm$	0.05	&	19.73	$\pm$	0.02	&	19.33	$\pm$	0.02	& 19.69	$\pm$	0.02 &	19.58	$\pm$	0.02	&	Sc20, Lo22	\\
182831.08$+$265037.8	&	Y2	&	24.47	$\pm$	 0.05		&	24.69	$\pm$	 0.05		&	24.32	$\pm$	 0.05		& 25.47 $\pm$	 0.05		&	24.71	$\pm$	 0.05		&	 R-GTO 	\\
200520.38$+$542433.9$^{**}$ 	&	T8	&	21.43	$\pm$	0.05	&	21.13	$\pm$	0.04	 &	21.29	$\pm$	0.05	& &			&	Ma13	\\
205628.90$+$145953.3	&	Y0	&	20.76	$\pm$	 0.05	&	20.82	$\pm$	 0.05		&	21.41	$\pm$	 0.05		& 23.02	$\pm$	0.25 &	21.65	$\pm$	 0.10		&	Be24	\\
210200.15$-$442919.5	&	T9	&	20.28	$\pm$	 0.05		&	19.84	$\pm$	 0.05		&	20.28	$\pm$	 0.05		& 21.75 $\pm$	 0.05	 &	20.41	$\pm$	 0.05		&	Be24	\\
214638.83$-$001038.7	&	T8.5	&			&	19.82	$\pm$	0.03	&	20.46	$\pm$	0.03	& 21.49 $\pm$	0.10 &	20.64	$\pm$	0.05	&	Bn09	\\
215949.54$-$480855.2 	&	T9	&	20.98	$\pm$	 0.05		&	20.49	$\pm$	 0.05		&	20.92	$\pm$	 0.05		& 22.60 $\pm$	 0.07	 &	21.82	$\pm$	 0.07		&	Be24	\\
221216.33$-$693121.6	&	T9	&	21.62	$\pm$	0.03	&	21.41	$\pm$	0.03	&			&	& 		&	Sc15	\\
222055.31$-$362817.4	&	Y0	&	22.45	$\pm$	0.07	&	22.25	$\pm$	0.03	&	22.64	$\pm$	0.04	& &				&	Sc15	\\
232519.54$-$410534.9	&	T9	&	21.45	$\pm$	0.12	&	21.18	$\pm$	0.03	&	21.53	$\pm$	0.10	& &	22.56	$\pm$	0.30	&	Ma13	\\
235402.79$+$024014.1 	&	Y1	&	24.06	$\pm$	 0.16		&	24.98	$\pm$	 0.18	&	25.02	$\pm$	0.20	& &				&	Be24	\\
\enddata
\tablenotetext{*} {Also known as COCONUTS-2b}
\tablenotetext{**} {\ Also known as Wolf 1130B}
\tablerefs{
Bg10 -- \citet{Burgasser_2010};
Bn08 -- \citet{Burningham_2008};
Bn09 -- \citet{Burningham_2009};
Bn10 -- \citet{Burningham_2010a};
Be23 -- \citet{Beiler_2023}; 
Be24 -- \citet{Beiler_2024}; 
Cu11 -- \citet{Cushing_2011}; 
Cu21 -- \citet{Cushing_2021};
De08 -- \citet{Delorme_2008};
Fa25 -- Faherty et al. 2025, Nature in press;
Ki11 -- \citet{Kirkpatrick_2011};
Le16 -- \citet{Leggett_2014};
Le16 -- \citet{Leggett_2016a}; 
Le17 -- \citet{Leggett_2017};
Lo22 -- \citet{Lodieu_2022}; 
Lu24 -- \citet{Luhman_2024};
Ma13 -- \citet{Mace_2013a}; 
R-GTO -- GTO 1189, PI = Roellig; 
Pi14 -- \citet{Pinfield_2014b};
Sc15 -- \citet{Schneider_2015};
Sc20 -- \citet{Schneider_2020};
Zh25 -- \citet{Zhang_2025_COCO}.
}
\end{deluxetable*}

\vskip -0.5in
Figure 1 illustrates the location of the {\it Roman} near-infared filters F106, F129, F158, F184, and F213. F106 and F129 are similar in bandpass to the MKO $Y$ and $J$ filters, although slightly broader. F158 is very similar to  MKO $H$,
and F213 is slightly bluer than MKO $K$. The F184 filter samples a region of strong absorption (Figure 1) and is not ideal for brown dwarf work; however the {\it Roman} High Latitude Imaging Survey is currently specified to use the F106, F129, F158 and F184 filters \citep{Troxel_2023} and so we include it here, as well as the F213 filter.  The F213 filter is  impacted by thermal noise \citep{Bahvin_2023}, however there are arguments in favor of its use for extragalactic work \citep{Gomez_2023}, as well as brown dwarf studies. 

Figure 12 can be compared to the MKO-equivalent diagram Figure 8. Note however that the MKO system uses Vega magnitudes while {\it Roman} uses AB. The redward shifts in various $YJH$ colors for the very metal-poor dwarfs 1534$-$10 and 1810$-$10 can be seen in MKO, {\it Euclid}, and {\it Roman} colors --- Figure 8, 9, and 12.  There is very little signal in the F184 filter and these magnitudes are faint and more noisy (Table 5). The F158 $-$ F213 color can be better measured than F158 $-$ F184 (Table 5 and Figure 12), and this color shows significant intrinsic spread. The models suggest a strong dependence on metallicity.

\begin{figure}
\vskip -0.5in
\hskip 0.6in
\includegraphics[angle=0, width = 6.0 in]{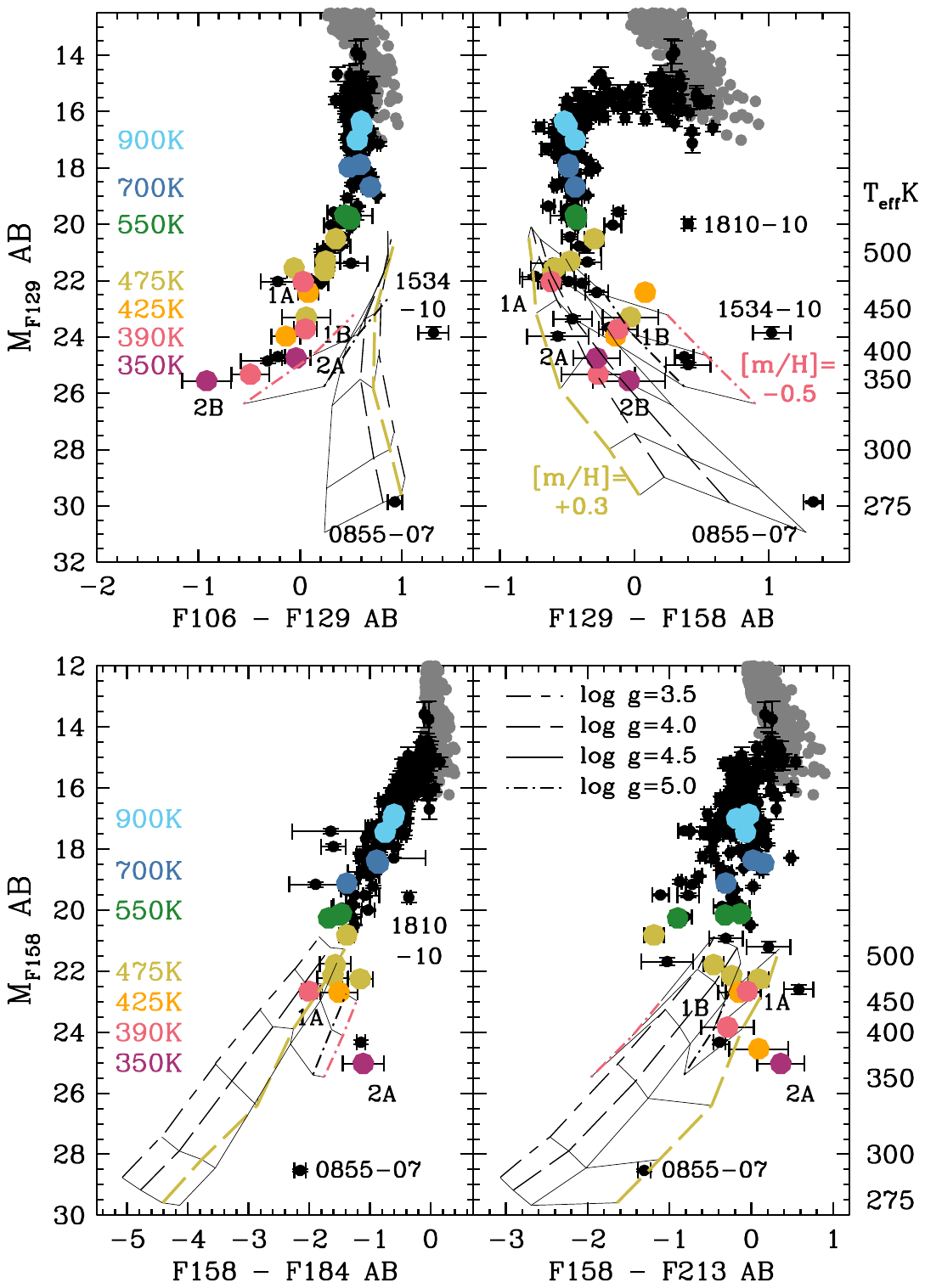}
\vskip -0.5in
\caption{{\it Roman} colors, note the very different $x$-axis scales.
Grey dots are M and L dwarfs, black dots are T and Y dwarfs. Colored symbols are dwarfs from \citet{Beiler_2024} where the color indicates $T_{\rm eff}$ value, shown in the legends in the left panels.
Red, yellow, and black almost vertical lines are ATMO 2020++ sequences with model parameters shown in the legends. Lines across the sequences are isotherms with  $T_{\rm eff}$ values indicated on the right axes. The offset in F106 $-$ F129 is due to the model spectra being too faint in the $Y$-band, see Section 4 and Figure 7.
The 275~K Y dwarf 0855$-$07 is identified, as are the metal-poor dwarfs 1534$-$10 and 1810$-$10.  
}
\end{figure}

\begin{figure}
\vskip -0.6in
\hskip 0.4in
\includegraphics[angle=0, width = 6 in]{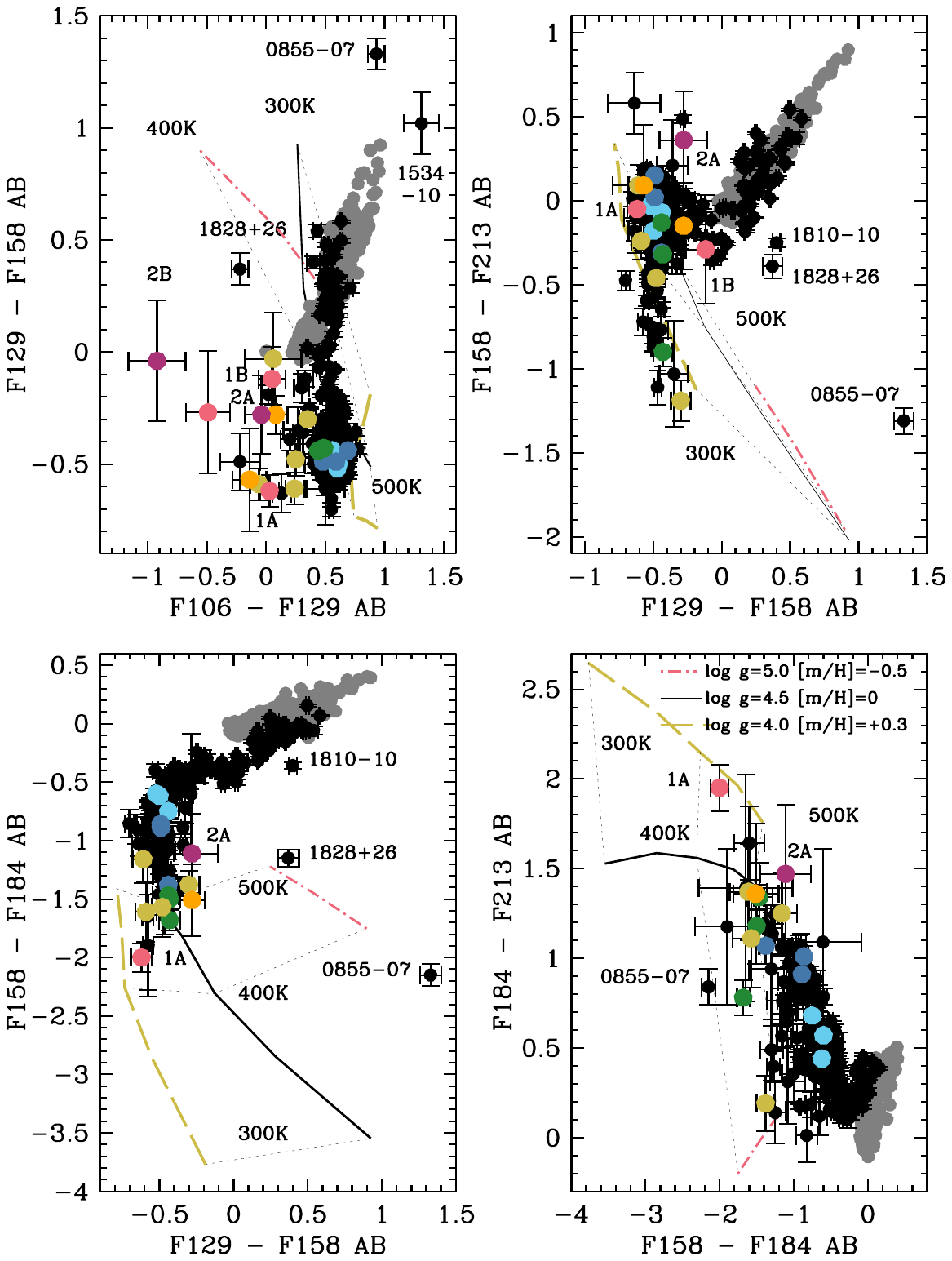}
\vskip -0.5in
\caption{{\it Roman} colors with symbols as in Figure 12.  Note the very different $x$-axis scales. Three model sequences are shown which sample the range of model colors in Figure 12. Parameters are given in the legend. The log $g = 4.0$ and 4.5 sequences span $300 \leq T_{\rm eff} \leq 500$, and the 
log $g = 5.0$ sequence spans $400 \leq T_{\rm eff} \leq 500$ (colder high gravity dwarfs are older than 15~Gyr \citep[and Figure 3]{Marley_2021}). Dotted lines join the 300, 400, and 500~K points across model sequences. The models are flux-deficient in the $Y$-band, resulting in a red offset in F106 $-$ F129. 
1828$+$26 is likely to be a high-gravity  unresolved binary \citep{Leggett_2025}. }
\end{figure}

The reference High Latitude Survey is expected to reach depths of around 26.9 mag AB in F106, F129, and F158, and 26.2 mag AB in F184 \citep{Troxel_2023}. Figure 12 shows that 350~K Y dwarfs at 10~pc will be detected in F106, F129, and F158. With $-2 \lesssim$ F158 $-$ F184 $\lesssim -1$, a detection in F184 will be challenging.
A lack of detection in F184 may nevertheless be scientifically valuable as an indicator of the presence of H$_2$O absorption. 

Figure 13 shows that it will be difficult to identify cold brown dwarfs in {\it Roman} data by near-infrared colors alone. The high latitude time domain surveys \citep{Rubin_2023} are likely to find the nearer and brighter brown dwarfs which will have significant proper motion \citep{Kirkpatrick_2010}. Classification will also be helped by multi-color detections. In the ideal case,  a low-resolution spectrum will be obtained \citep{Wang_2022}.

\clearpage
\section{Conclusions}

The mid-infrared wavelength region is critical for understanding cold brown dwarfs. We find that $M_{W2}$ can be used as a proxy for $T_{\rm eff}$ and we give an $M_{W2}$:$T_{\rm eff}$ polynomial relationship that can be used for single late-T and Y dwarfs with metallicities and ages typical of the local field (Figure 4). We find that the strength of the CO and CO$_2$ absorption features at $4 < \lambda~\mu$m $< 5$ constrain gravity and metallicity for Y dwarfs (Figure 5), and we use {\it JWST} spectra across this wavelength range, together with measured distances and evolutionary models, to estimate  $T_{\rm eff}$, log $g$, [m/H], mass and age for five Y dwarfs (Table 2). The ATMO 2020++ PH$_3$-free synthetic spectra fits the $0.9 < \lambda~\mu$m $< 15$ energy distributions of these dwarfs well, with the following exceptions:
\begin{itemize}
    \item For the apparently young low-gravity metal-rich 1047$+$54 (Table 2), the pressure-temperature (P-T) relationship needs to be adjusted so that the lower atmosphere is warmer and the upper atmosphere colder. 
    \item For the very cold (275~K) 0855$-$07, the models calculate overly strong NH$_3$ absorption at $9 < \lambda~\mu$m $< 15$, suggesting that significant changes to the atmospheric chemistry are needed at cold temperatures.
    \item The models calculate too little flux at $2 < \lambda~\mu$m $< 4$ suggesting that the top of the atmosphere ($P \approx 0.1$~bar) needs to be warmer.
    \item The models calculate too little flux in the $Y$-band, at $\lambda \approx 1~\mu$m. The models also show a spurious emission feature at $\lambda \approx 0.85~\mu$m for the very cold 0855$-$07. Modelling this wavelength region is complex because:
    \begin{itemize}
        \item the treatment of the red wing of the K~I resonance absorption is not well understood.
        \item The abundance of K~I is not well constrained due to grain formation.
        \item Currently the ATMO 2020++ models do not extend the H$_2$ opacity to wavelengths shorter than $1.0~\mu$m, and do not extend the CH$_4$ or NH$_3$ opacity shorter than  $0.8~\mu$m. 
    \end{itemize}    
\end{itemize}


Although the ATMO 2020++ models show systematic discrepancies in the $Y$- and $K$-bands those regions contribute $\lesssim 1\%$ of the total energy of a 400~K Y dwarf (based on the {\it JWST} spectral energy distribution for 1405$+$55). The $J$- and $H$-band flux peaks are generally well reproduced (Figure 7). 

We calculated new synthesized near-infrared colors for late-T and Y dwarfs in the MKO, {\it Euclid}, and {\it Roman} photometric systems (Sections 3, 5 and 6). We also synthesized  {\it Euclid} $I_E$ magnitudes for 15 T and Y dwarfs. Trends can be seen in the near-infrared color diagrams however the absolute brightness shows a larger range with temperature than the mid-infrared luminosity, and the colors suggest more subtle or complex dependencies on temperature, metallicity and gravity (Figure 8).

The brown dwarfs with [m/H] $< -1.5$ stand out in near-infrared color diagrams --- they are red in $J - H$, $Y_E - J_E$, $J_E - H_E$, F106 $-$ F129, F129 $-$ F158, F158 $-$ F184 (Figures 8, 9, 12). The coldest known brown dwarf, the 275~K Y dwarf 0855$-$07, mostly lies on an extensions of the warmer color sequences in the color-color diagrams, but at shorter wavelengths the color trend reverses: $Y - J$, $Y_E - J_E$, $I_E - H_E$, $I_E - Y_E$  (Figures 8, 9, 12).


{\it Euclid} and {\it Roman} will be sensitive to Y dwarfs. For objects without parallaxes it will be difficult to identify them using colors alone (Figures 10 and 13).  As the objects will of necessity be nearby, proper motion can be used as a low-luminosity indicator. For {\it Euclid}, cold brown dwarfs will be very red in $I_E - H_E$ (Figure 11). With regard to $K$-band filter choice for {\it Roman}, brown dwarfs are much brighter in the F213 bandpass than the F184 bandpass, however  F184 could be used as a ``dropout'' filter, indicating the presence of H$_2$O absorption.

\begin{acknowledgements}

MWP acknowledges support from UKRI STFC AGP grant ST/W001209/1 and the University of Edinburgh School of Physics \& Astronomy.

\end{acknowledgements}

\appendix

\section{MKO, {\it Spitzer}, and {\it WISE} Photometry Compilation}

Table 6 presents a compilation of observational MKO, {\it Spitzer}, and {\it WISE} photometry. 

We include new data for two mid-T dwarfs, which, although not the 
primary target type of interest for this work, we consider to be valuable: 
\begin{itemize}
    \item For the T6.5 brown dwarf GJ 229B we have synthesized $YJHK$.  We used the red and near-infrared spectra published by \citet{Geballe_1996} and
\citet{Oppenheimer_1998}, together with {\it HST} and ground-based photometry \citep{Golimowski_1998, Leggett_1999}, to produce a flux calibrated spectrum from which we determined improved colors.  The system has 
recently been found to be a close binary system \citep{Xuan_2024}. 
\item For the metal-poor T6.5 brown dwarf ULAS 131610.28 $+$075553.0 we have determined $H$ from the UKIDSS LAS image via aperture photometry. $Y$ and $J$ were presented in \citet{Burningham_2014}.
 \end{itemize}

Tables 7 and 8 present the ATMO 2020++ {\it Euclid} and {\it Roman} colors for the models used in Figures 9, 10, 12, and 13.

\startlongtable
\begin{deluxetable*}{lrrcrrrrrrrllll}
\setlength{\tabcolsep}{1pt}
\tabletypesize{\scriptsize}
\tablecaption{Compilation of Photometric Measurements for T- and Y-Type Brown Dwarfs}
\tablehead{
\\[0.005in]
\colhead{Survey} 
&	\multicolumn{2}{c}{Discovery R.A. Decl.} 
&	\colhead{Sp.} 
&	\colhead{M $-$ m}	
& \colhead{$Y$} 
&	\colhead{$J$} 
&	\colhead{W2}	
& \colhead{$e_Y$}	
& \colhead{$e_J$} 
&  \colhead{$e_{\rm W2}$} 
&	\multicolumn{4}{c}{References}\\
\colhead{Name}	
& \colhead{hhmmss.ss} 
& \colhead{$\pm$ddmmss.s}	
& \colhead{Type} 
&	\multicolumn{7}{c}{mag} 
& \colhead{Discovery} 
&	\colhead{Sp. Type}	
& \colhead{Parallax}
& \colhead{Near-IR}	
}
\startdata
CWISEP	& 000229.93	& $+$635217.0 &	7.5	 &	 &	& & 15.73 & & & 
0.05 & Meisner\_2020b	& Meisner\_2020b	&  &	  \\
2MASS  &   000430.65 & $-$260359.6 & 2.0 & $-2.40$  & 17.28 & 16.18 & 14.15 & 0.06 & 0.02 & 0.02 & Schneider\_2016 & Greco\_2019  & Best\_2020 &  Best\_2021  \\
WISE   &   000517.48 & $+$373720.5 & 9.0 & 0.52  & 18.48  & 17.59 & 13.30 & 0.02 & 0.02 & 0.01 & Mace\_2013a  & Mace\_2013a   & Kirkpatrick\_2019   & Leggett\_2015     \\                          
CWISEP &   001146.07 & $-$471306.8 & 8.5 &  &  & 19.28 & 15.99  & & 0.07 & 0.05 &  Meisner\_2020a   &  Meisner\_2020a    &  &   VISTA\_VHS   \\
2MASS   &  001322.29 & $-$114300.6 & 3.0 & $-1.97$ & & 16.05 & 14.36 & & 0.02 & 0.02 & Kellogg\_2017 &   Kellogg\_2017  & Best\_2020  & Best\_2021               \enddata
\tablecomments{Table 6 is published in its entirety in the machine-readable format. A portion is shown here for guidance regarding its form and content.}
\vskip -0.05in
\tablerefs{
2MASS -- \citet{Skrutskie_2006};
Albert${\_}$2011 -- \citet{Albert_2011};
Artigau${\_}$2010 -- \citet{Artigau_2010};
Bardalez${\_}$2020 -- \citet{Bardalez_2020};
Beichman${\_}$2014 -- \citet{Beichman_2014}; 
Beiler${\_}$2024 -- \citet{Beiler_2024};
Best${\_}$2015 -- \citet{Best_2015};
Best${\_}$2020 -- \citet{Best_2020};
Best${\_}$2021 -- \citet{Best_2021};
Best${\_}$2024 -- \citet{Best_2024};
Bihain${\_}$2013  -- \citet{Bihain_2013};
Brooks${\_}$2022 -- \citet{Brooks_2022};
Burgasser${\_}$1999 -- \citet{Burgasser_1999};
Burgasser${\_}$2000 -- \citet{Burgasser_2000};
Burgasser${\_}$2002 -- \citet{Burgasser_2002};
Burgasser${\_}$2003 -- \citet{Burgasser_2003};
Burgasser${\_}$2004 -- \citet{Burgasser_2004};
Burgasser${\_}$2006 -- \citet{Burgasser_2006a};
Burgasser${\_}$2008 -- \citet{Burgasser_2008};
Burgasser${\_}$2010 -- \citet{Burgasser_2010}
Burgasser${\_}$2012 -- \citet{Burgasser_2012}
Burningham${\_}$2008 -- \citet{Burningham_2008};
Burningham${\_}$2009 -- \citet{Burningham_2009};
Burningham${\_}$2010a -- \citet{Burningham_2010a};
Burningham${\_}$2010b -- \citet{Burningham_2010b};
Burningham${\_}$2011 -- \citet{Burningham_2011};
Burningham${\_}$2013 -- \citet{Burningham_2013};
Burningham${\_}$2014 -- \citet{Burningham_2014};
Chiu${\_}$2006 -- \cite{Chiu_2006};
Cushing${\_}$2011 -- \citet{Cushing_2011};
Cushing${\_}$2014 -- \citet{Cushing_2014};
Cushing${\_}$2016 -- \citet{Cushing_2016};
Deacon${\_}$2011   -- \citet{Deacon_2011}; 
Deacon${\_}$2012  -- \citet{Deacon_2012};
Delorme${\_}$2008 -- \citet{Delorme_2008};
Delorme${\_}$2010 -- \citet{Delorme_2010};
Dupuy${\_}$2012 -- \citet{Dupuy_2012};
Dupuy${\_}$2013 -- \citet{Dupuy_2013};
Dupuy${\_}$2015 -- \citet{Dupuy_2015};
Dupuy${\_}$2017 -- \citet{Dupuy_2017};
Ellis${\_}$2005 -- \citet{Ellis_2005};
Faherty${\_}$2012 -- \citet{Faherty_2012};
Faherty${\_}$2020 -- \citet{Faherty_2020};
Faherty${\_}$2025 -- Faherty et al. 2025, Nature in press;
Fontanive${\_}$2021   -- \citet{Fontanive_2021};        Fontanive${\_}$2025   -- \citet{Fontanive_2025};   
Gaia -- \citet{GAIA};
Geballe${\_}$2001  -- \citet{Geballe_2001};
Geballe${\_}$2002  -- \citet{Geballe_2002};
Gelino${\_}$2011 -- \citet{Gelino_2011};
Goldman${\_}$2010 -- \citet{Goldman_2010};
Golimowski${\_}$2004 -- \citet{Golimowski_2004};
Greco${\_}$2019 -- \citet{Greco_2019};
Griffith${\_}$2012 -- \citet{Griffith_2012};
Kellogg${\_}$2017  -- \citet{Kellogg_2017};
Kirkpatrick${\_}$2011 -- \citet{Kirkpatrick_2011};
Kirkpatrick${\_}$2012 -- \citet{Kirkpatrick_2012};
Kirkpatrick${\_}$2013  -- \citet{Kirkpatrick_2013};
Kirkpatrick${\_}$2019  -- \citet{Kirkpatrick_2019};
Kirkpatrick${\_}$2021a  -- \citet{Kirkpatrick_2021a};
Kirkpatrick${\_}$2021b  -- \citet{Kirkpatrick_2021b};
Knapp${\_}$2004 -- \citet{Knapp_2004};
Kota${\_}$2022 -- \citet{Kota_2022};
Leggett${\_}$2002 -- \citet{Leggett_2002};
Leggett${\_}$2009 -- \citet{Leggett_2009};
Leggett${\_}$2010 -- \citet{Leggett_2010};
Leggett${\_}$2012 -- \citet{Leggett_2012};
Leggett${\_}$2013 -- \citet{Leggett_2013};
Leggett${\_}$2014 -- \citet{Leggett_2014};
Leggett${\_}$2015 -- \citet{Leggett_2015};
Leggett${\_}$2017 -- \citet{Leggett_2017};
Leggett${\_}$2019 -- \citet{Leggett_2019};
Leggett${\_}$2021 -- \citet{Leggett_2021};
L25, Leggett${\_}$2025${\_}$UKIDSS${\_}$LAS, Leggett${\_}$2025${\_}$VISTA${\_}$VVX --  \citet{Leggett_2025};
Liu${\_}$2007 -- \citet{Liu_2007};
Liu${\_}$2011 -- \citet{Liu_2011};
Liu${\_}$2012 -- \citet{Liu_2012};
Lodieu${\_}$2007 -- \citet{Lodieu_2007};
Lodieu${\_}$2009 -- \citet{Lodieu_2009};
Lodieu${\_}$2012 -- \citet{Lodieu_2012};
Lodieu${\_}$2022 -- \citet{Lodieu_2022};
Looper${\_}$2007 -- \citet{Looper_2007};
Lucas${\_}$2010 -- \citet{Lucas_2010};
Luhman${\_}$2011 -- \citet{Luhman_2011};
Luhman${\_}$2012 -- \citet{Luhman_2012};
Luhman${\_}$2014 -- \citet{Luhman_2014};
Mace${\_}$2013a -- \citet{Mace_2013a}; 
Mace${\_}$2013b -- \citet{Mace_2013b}; 
Mainzer${\_}$2011 -- \citet{Mainzer_2011};
Manjavacas${\_}$2013 -- \citet{Manjavacas_2013};
Marocco${\_}$2010 -- \cite{Marocco_2010};
Marocco${\_}$2015 -- \cite{Marocco_2015};
Marocco${\_}$2019 -- \cite{Marocco_2019};
Martin${\_}$2018 -- \citet{Martin_2018};
Meisner${\_}$2020a -- \citet{Meisner_2020a};
Meisner${\_}$2020b -- \citet{Meisner_2020b};
Meisner${\_}$2021 -- \citet{Meisner_2021};
Meisner${\_}$2023 -- \citet{Meisner_2023};
Meisner${\_}$2024RN -- \citet{Meisner_2024};
Mugrauer${\_}$2006  -- \citet{Mugrauer_2006};
Murray${\_}$2011 -- \citet{Murray_2011};
Nakajima${\_}$1995  -- \citet{Nakajima_1995};
Patten${\_}$2006 -- \citet{Patten_2006};
Pinfield${\_}$Gromadzki${\_}$2014 -- Pinfield, P. and Gromadzki, M. private communication 2014;
Pinfield${\_}$2008 -- \citet{Pinfield_2008};
Pinfield${\_}$2012 -- \citet{Pinfield_2012};
Pinfield${\_}$2014a -- \citet{Pinfield_2014a};
Pinfield${\_}$2014b -- \citet{Pinfield_2014b};
Robbins${\_}$2023     -- \citet{Robbins_2023};   
Rothermich${\_}$2024    -- \citet{Rothermich_2024};   
Schneider${\_}$2015     -- \citet{Schneider_2015}; 
Schneider${\_}$2016  -- \citet{Schneider_2016};
Schneider${\_}$2020      -- \citet{Schneider_2020}; 
Schneider${\_}$2021  -- \citet{Schneider_2021};
Scholz${\_}$2010a -- \citet{Scholz_2010a};
Scholz${\_}$2010b -- \citet{Scholz_2010b};
Scholz${\_}$2011 -- \citet{Scholz_2011};
Scholz${\_}$2012 -- \citet{Scholz_2012};
Sheppard${\_}$2009 -- \citet{Sheppard_2009};
Smart${\_}$2010 -- \citet{Smart_2010};
Smart${\_}$2013 -- \citet{Smart_2013};
Strauss${\_}$1999 -- \cite{Strauss_1999};
Subasavage${\_}$2009 -- \citet{Subasavage_2009};
Thompson${\_}$2013 -- \citet{Thompson_2013};
Tinney${\_}$2003 -- \citet{Tinney_2003};
Tinney${\_}$2005 -- \citet{Tinney_2005};
Tinney${\_}$2012 -- \citet{Tinney_2012};
Tinney${\_}$2014 -- \citet{Tinney_2014};
Tinney${\_}$2018 -- \citet{Tinney_2018};
Tsvetanov${\_}$2000 -- \citet{Tsvetanov_2000};
UKIDSS -- \citet{Lawrence_2007};
VISTA -- \citet{Sutherland_2015};
Vrba${\_}$2004 -- \citet{Vrba_2004};
Warren${\_}$2007 -- \citet{Warren_2007};
Wright${\_}$2013 -- \citet{Wright_2013};
Zhang${\_}$2020   -- \citet{Zhang_2020};                Zhang${\_}$2024 -- \citet{Zhang_2024};
Zhang${\_}$2025 -- \citet{Zhang_2025_2217}.
}      
\end{deluxetable*}

\begin{deluxetable*}{cccccccccccccc}
\tabletypesize{\small}
\tablecaption{ATMO 2020++ PH$_3$-Free {\it Euclid} Near-Infrared AB mags for $D = 10~$pc and $R = 0.1~R_{\odot}$ }
\tablehead{
\colhead{$T_{\rm eff}$} & 
\colhead{log $g$} &
\colhead{[m/H]} &
\colhead{$Y_E$} & 
\colhead{$J_E$} &
\colhead{$H_E$} & 
\colhead{[m/H]} &
\colhead{$Y_E$} & 
\colhead{$J_E$} &
\colhead{$H_E$} & 
\colhead{[m/H]} &
\colhead{$Y_E$} & 
\colhead{$J_E$} &
\colhead{$H_E$}  
}
\startdata
275 & 3.5 & 0.0  &  30.30 & 29.37 & 29.18  &  0.3  & 30.14	 & 29.15 & 29.55    & -0.5  & 30.39	 & 30.07 &  29.01 \\
300 & 3.5 & 0.0  &  28.70 & 27.85 & 27.91  &  0.3  & 28.33	 & 27.46 & 28.04    & -0.5  & 28.79	 & 28.45 &  27.75 \\
350 & 3.5 & 0.0  &  25.85 & 25.12 & 25.57  &  0.3  & 25.91	 & 25.12 & 25.90    & -0.5  & 26.28	 & 25.80 & 	25.65 \\
400 & 3.5 & 0.0  &  23.91 & 23.13 & 23.82  &  0.3  & 23.93	 & 23.16 & 24.03    & -0.5  & 24.21	 & 23.61 & 	23.91 \\
450 & 3.5 & 0.0  &  22.56 & 21.70 & 22.51  &  0.3  & 22.55	 & 21.7	 & 22.65    & -0.5  & 22.71  & 22.02 & 	22.59 \\
500 & 3.5 & 0.0  &  21.44 & 20.59 & 21.46  &  0.3  & 21.51	 & 20.6	 & 21.60    & -0.5  & 21.47	 & 20.76 & 	21.51 \\
275 & 4.0 & 0.0  & 30.90 & 	30.18 & 29.52 &  0.3  & 30.82 & 	29.95 & 	29.91 & -0.5  & 29.90 & 	30.54 & 	29.17 \\
300 & 4.0 & 0.0  & 29.27 & 	28.63 & 28.27 &  0.3  & 29.12 & 	28.35 & 	28.53 & -0.5  & 28.65 & 	29.03 & 	28.01 \\
350 & 4.0 & 0.0  & 26.36 & 	25.84 & 25.99 &  0.3  & 26.25 & 	25.63 & 	26.17 & -0.5  & 26.60 & 	26.51 & 	26.05 \\
400 & 4.0 & 0.0  & 24.32 & 	23.72 & 24.23 &  0.3  & 24.21 & 	23.53 & 	24.26 & -0.5  & 24.67 & 	24.30 & 	24.34 \\
450 & 4.0 & 0.0  & 22.91 & 	22.12 & 22.80 &  0.3  & 22.87 & 	22.05 & 	22.81 & -0.5  & 23.18 & 	22.61 & 	22.99 \\
500 & 4.0 & 0.0  & 21.77 & 	20.90 & 21.66 &  0.3  & 21.78 & 	20.87 & 	21.66 & -0.5  & 21.92 & 	21.26 & 	21.86 \\
275 & 4.5 & 0.0  & 31.17 & 	30.98 & 29.79 &  0.3  & 31.73 & 30.95 & 30.28 & -0.5  & 29.04 & 30.89 & 29.38 \\
300 & 4.5 & 0.0  & 29.63 & 	29.42 & 28.58 &  0.3  & 29.64 & 29.01 & 28.73 & -0.5  & 28.04 & 29.42 & 28.21 \\
350 & 4.5 & 0.0  & 26.80 & 	26.55 & 26.32 &  0.3  & 26.72 & 26.23 & 26.39 & -0.5  & 26.43 & 26.96 & 26.27 \\
400 & 4.5 & 0.0  & 24.76 & 	24.35 & 24.54 &  0.3  & 24.56 & 24.04 & 24.48 & -0.5  & 25.05 & 24.96 & 24.69 \\
450 & 4.5 & 0.0  & 23.38 & 	22.71 & 23.12 &  0.3  & 23.18 & 22.41 & 22.98 & -0.5  & 23.67 & 23.21 & 23.31 \\
500 & 4.5 & 0.0  & 22.16 & 	21.30 & 21.88 &  0.3  & 22.12 & 21.13 & 21.75 & -0.5  & 22.48 & 21.84 & 22.17 \\
275 & 5.0 & 0.0  & 30.52 & 31.39 & 29.78 &  0.3  & 31.62 & 31.20 & 29.97 & -0.5  & 28.17 & 31.21 & 29.78 \\
300 & 5.0 & 0.0  & 29.19 & 29.79 & 28.52 &  0.3  & 29.97 & 29.55 & 28.66 & -0.5  & 27.46 & 29.82 & 28.57 \\
350 & 5.0 & 0.0  & 26.98 & 27.10 & 26.41 &  0.3  & 27.20 & 26.79 & 26.42 & -0.5  & 26.15 & 27.41 & 26.53 \\
400 & 5.0 & 0.0  & 25.19 & 24.96 & 24.70 &  0.3  & 24.96 & 24.51 & 24.57 & -0.5  & 25.03 & 25.38 & 24.83 \\
450 & 5.0 & 0.0  & 23.64 & 23.09 & 23.19 &  0.3  & 23.39 & 22.72 & 23.04 & -0.5  & 23.96 & 23.70 & 23.45 \\
500 & 5.0 & 0.0  & 22.55 & 21.69 & 22.00 &  0.3  & 22.33 & 21.35 & 21.82 & -0.5  & 22.89 & 22.28 & 22.28 \\
\enddata
\end{deluxetable*}

\startlongtable
\begin{deluxetable*}{cccccccccccccccccccc}
\setlength{\tabcolsep}{2pt}
\tabletypesize{\scriptsize}
\tablecaption{ATMO 2020++ PH$_3$-Free {\it Roman} Near-Infrared AB mags for $D = 10~$pc and $R = 0.1~R_{\odot}$ }
\tablehead{
\\[0.005in]
\colhead{$T_{\rm eff}$} & 
\colhead{log $g$} &
\colhead{[m/H]} &
\colhead{F106} & 
\colhead{F129} &
\colhead{F158} & 
\colhead{F184} & 
\colhead{F213} &
\colhead{[m/H]} &
\colhead{F106} & 
\colhead{F129} &
\colhead{F158} & 
\colhead{F184} & 
\colhead{F213} &
\colhead{[m/H]} &
\colhead{F106} & 
\colhead{F129} &
\colhead{F158} & 
\colhead{F184} & 
\colhead{F213} 
}
\startdata
275 & 3.5 & 0.0  & 30.31  & 29.29  & 29.07  & 34.15  & 32.14 &  0.3  & 30.15  & 29.04  & 29.44  & 33.88  & 31.19 & -0.5  & 30.41  & 30.13  & 28.88  & 33.60  & 33.86 \\
300 & 3.5 & 0.0  & 28.72  & 27.79  & 27.79  & 32.23  & 30.25 &  0.3  & 28.34  & 27.36  & 27.94  & 31.90  & 29.32 & -0.5  & 28.81  & 28.48  & 27.62  & 31.94  & 31.86 \\
350 & 3.5 & 0.0  & 25.87  & 25.10  & 25.47  & 29.02  & 27.10 &  0.3  & 25.93  & 25.06  & 25.83  & 28.79  & 26.49 & -0.5  & 26.31  & 25.86  & 25.53  & 29.18  & 28.51 \\
400 & 3.5 & 0.0  & 23.94  & 23.15  & 23.73  & 26.53  & 24.78 &  0.3  & 23.96  & 23.15  & 23.98  & 26.34  & 24.25 & -0.5  & 24.25  & 23.68  & 23.79  & 26.84  & 26.01 \\
450 & 3.5 & 0.0  & 22.60  & 21.73  & 22.44  & 24.68  & 23.10 &  0.3  & 22.58  & 21.75  & 22.61  & 24.55  & 22.70 & -0.5  & 22.76  & 22.10  & 22.47  & 24.97  & 24.11 \\
500 & 3.5 & 0.0  & 21.49  & 20.63  & 21.40  & 23.24  & 21.86 &  0.3  & 21.56  & 20.70  & 21.58  & 23.11  & 21.49 & -0.5  & 21.52  & 20.84  & 21.38  & 23.51  & 22.70 \\
275 & 4.0 & 0.0  & 30.92 & 30.10 & 29.40 & 34.18 & 32.30 &  0.3  & 30.84 & 29.84 & 29.81 & 34.22 & 31.45 & -0.5  & 29.93 & 30.67 & 29.04 & 33.41 & 33.90 \\
300 & 4.0 & 0.0  & 29.29 & 28.57 & 28.16 & 32.30 & 30.39 &  0.3  & 29.13 & 28.25 & 28.44 & 32.21 & 29.56 & -0.5  & 28.67 & 29.12 & 27.89 & 31.76 & 31.80 \\
350 & 4.0 & 0.0  & 26.39 & 25.80 & 25.90 & 29.17 & 27.31 &  0.3  & 26.27 & 25.55 & 26.11 & 28.96 & 26.59 & -0.5  & 26.63 & 26.58 & 25.94 & 29.07 & 28.63 \\
400 & 4.0 & 0.0  & 24.35 & 23.71 & 24.16 & 26.75 & 25.04 &  0.3  & 24.23 & 23.49 & 24.23 & 26.48 & 24.34 & -0.5  & 24.71 & 24.36 & 24.23 & 26.86 & 26.25 \\
450 & 4.0 & 0.0  & 22.95 & 22.13 & 22.75 & 24.82 & 23.19 &  0.3  & 22.91 & 22.05 & 22.81 & 24.58 & 22.61 & -0.5  & 23.23 & 22.67 & 22.89 & 25.07 & 24.36 \\
500 & 4.0 & 0.0  & 21.83 & 20.93 & 21.63 & 23.28 & 21.81 &  0.3  & 21.83 & 20.89 & 21.67 & 23.09 & 21.33 & -0.5  & 21.98 & 21.33 & 21.75 & 23.63 & 22.90 \\
275 & 4.5 & 0.0  & 31.19 & 330.96 & 329.68 & 333.80 & 332.36 &  0.3  & 31.74 & 330.86 & 330.17 & 334.08 & 331.46 & -0.5  & 29.07 & 31.22 & 	29.26 & 	33.06 & 	33.85 \\
300 & 4.5 & 0.0  & 29.66 & 329.39 & 328.47 & 332.01 & 330.49 &  0.3  & 29.65 & 328.92 & 328.64 & 332.17 & 329.63 & -0.5  & 28.07 & 29.65 & 	28.09 & 	31.44 & 	31.89 \\
350 & 4.5 & 0.0  & 26.83 & 226.52 & 226.23 & 229.08 & 227.49 &  0.3  & 26.74 & 226.16 & 226.32 & 229.08 & 226.73 & -0.5  & 26.45 & 27.09 & 	26.15 & 	28.78 & 	28.78 \\
400 & 4.5 & 0.0  & 24.80 & 224.34 & 224.47 & 226.77 & 225.21 &  0.3  & 24.59 & 224.00 & 224.44 & 226.66 & 224.50 & -0.5  & 25.07 & 25.05 & 	24.58 & 	26.68 & 	26.43 \\
450 & 4.5 & 0.0  & 23.43 & 222.72 & 223.07 & 224.89 & 223.39 &  0.3  & 23.22 & 222.41 & 222.97 & 224.73 & 222.78 & -0.5  & 23.71 & 23.28 & 	23.21 & 	24.94 & 	24.55 \\
500 & 4.5 & 0.0  & 22.22 & 221.34 & 221.85 & 223.39 & 221.98 &  0.3  & 22.18 & 221.17 & 221.76 & 223.19 & 221.41 & -0.5  & 22.54 & 21.91 & 	22.08 & 	23.52 & 	23.05 \\
275 & 5.0 & 0.0  & 30.56 & 	31.53 & 	29.66 & 	33.26 & 	32.56 &  0.3  & 31.65 & 	31.17 & 	29.86 & 	33.62 & 	31.55 &  -0.5  & 28.20 & 	31.88 & 	29.65 & 	32.68 & 	34.00 \\
300 & 5.0 & 0.0  & 29.23 & 	29.87 & 	28.41 & 	31.52 & 	30.65 &  0.3  & 29.99 & 	29.51 & 	28.56 & 	31.81 & 	29.72 &  -0.5  & 27.49 & 	30.34 & 	28.44 & 	31.11 & 	32.11 \\
350 & 5.0 & 0.0  & 27.01 & 	27.13 & 	26.31 & 	28.73 & 	27.65 &  0.3  & 27.23 & 	26.76 & 	26.34 & 	28.85 & 	26.84 &  -0.5  & 26.17 & 	27.74 & 	26.41 & 	28.54 & 	29.08 \\
400 & 5.0 & 0.0  & 25.23 & 	24.99 & 	24.63 & 	26.56 & 	25.45 &  0.3  & 24.99 & 	24.49 & 	24.51 & 	26.60 & 	24.69 &  -0.5  & 25.05 & 	25.61 & 	24.71 & 	26.46 & 	26.66 \\
450 & 5.0 & 0.0  & 23.69 & 	23.12 & 	23.13 & 	24.77 & 	23.60 &  0.3  & 23.44 & 	22.73 & 	23.00 & 	24.78 & 	22.95 &  -0.5  & 23.99 & 	23.87 & 	23.33 & 	24.78 & 	24.79 \\
500 & 5.0 & 0.0  & 22.63 & 	21.74 & 	21.95 & 	23.35 & 	22.23 &  0.3  & 22.40 & 	21.39 & 	21.80 & 	23.30 & 	21.62 &  -0.5  & 22.94 & 	22.42 & 	22.17 & 	23.39 & 	23.27 \\
\enddata
\end{deluxetable*}

\clearpage
\bibliography{NearIR_2025}{}
\bibliographystyle{aasjournal}

\end{document}